\documentclass[twocolumn]{aastex631}

\newcommand{\pcm}{\,$\mathrm{cm^{-2}}$ }
\newcommand{\lum}{\,$\mathrm{erg~s^{-1}}$}

\newcommand{\kev}{\,$\mathrm{keV}$}
\newcommand{\kpc}{\,$\mathrm{kpc}$}
\newcommand{\ksec}{\,$\mathrm{ksec}$}

\newcommand{\redchi}{\,$\mathrm{\chi_{\nu}^{2}}$ }
\newcommand{\chisq}{\,$\mathrm{\chi^{2}}$ }
\newcommand{\xspec}{\texttt{Xspec} }
\newcommand{\bwcycl}{\texttt{bwcycl}}
\newcommand{\gabs}{\texttt{gabs}}
\newcommand{\gauss}{\texttt{Gaussian}}
\newcommand{\cpl}{\texttt{cutoffpl}}
\newcommand{\powerlaw}{\texttt{powerlaw}}
\newcommand{\tbabs}{\texttt{tbabs}}
\newcommand{\phabs}{\texttt{phabs}}
\newcommand{\constant}{\texttt{constant}}
\newcommand{\tbpcf}{\texttt{tbpcf}}
\newcommand{\hcut}{\texttt{highecut}}
\newcommand{\bbody}{\texttt{bbody}}
\newcommand{\exo}{EXO~2030+375\xspace}
\newcommand{\hxmt}{\textsl{Insight-HXMT}\xspace}

\newcommand{\integral}{\textsl{INTEGRAL}\xspace}

\newcommand{\Tmound}{T_{\rm th}}

\newcommand{\vmound}{v_{\rm th}}

\newcommand{\taumound}{\tau_{\rm th}}

\newcommand{\taumax}{\tau_{\rm max}}

\newcommand{\tautrap}{\tau_{\rm trap}}

\def\sig{\sigma_{_{\rm T}}}
\def\sigpar{\sigma_\|}

\def\sigbar{\overline\sigma}

\usepackage{xspace}
\usepackage{graphicx}
\usepackage[outdir=./]{epstopdf}
\usepackage{graphicx}
\usepackage{amsmath}
\usepackage{multirow}

\begin{document}
\title{Evidence for a cyclotron absorption line and spectral transition in EXO 2030+375 during 2021 giant outburst}

\author{Wen Yang}
\affiliation{Department of Astronomy, School of Physics and Technology, Wuhan University, Wuhan 430072, China}
\author{Wei Wang}
\altaffiliation{Email address: wangwei2017@whu.edu.cn}
\affiliation{Department of Astronomy, School of Physics and Technology, Wuhan University, Wuhan 430072, China}
\author{Prahlad R. Epili}
\affiliation{Department of Astronomy, School of Physics and Technology, Wuhan University, Wuhan 430072, China}
 
\begin{abstract}
Based on Insight-HXMT observations of EXO 2030+375 during its 2021 giant outburst, we report the analysis of pulse variations and the broadband X-ray spectrum, and find the presence of a potential cyclotron resonant scattering feature (CRSF) with the fundamental line at $\sim47$ keV from both average spectra and phase-resolved spectroscopy. During the outburst, the source reached an X-ray luminosity of $\sim 10^{38}$ erg cm$^{-2}$ s$^{-1}$ from 2 – 105 keV at a distance of 7.1 kpc. 
The X-ray pulsar at the spin period of $41.27\pm 0.61$ seconds exhibits complex timing and spectral variations with both energy and luminosity during the outburst. The shapes of the pulses profiles show the single main peak above $\sim 20$ keV, while appear to exhibit multi-peak patterns in low energy bands, and the transition of the 10 -- 20 keV pulse profiles from multi-peak to single-peak is observed at $\sim 0.8\times10^{38}$ erg cm$^{-2}$ s$^{-1}$, which suggests the evolution from the subcritical luminosity (pencil-beam dominated) to supercritical luminosity 
(fan-beam dominated) regimes. A dip structure before the energy of the cyclotron resonant scattering features is found in the pulse fraction–energy relation near the peak luminosity. A detailed analysis of spectral parameters showed that the power-law photon index exhibits three distinct trends as luminosity increases, and these changes also signify a spectral transition from sub-critical to super-critical regimes. The critical luminosity infers the magnetic field of $\sim(4.8-6.0)\times 10^{12}$ G, which supports the presence of the cyclotron line at $\sim$47 keV. A Comptonization model applied for the broad X-ray spectra during the outburst also suggests the surface magnetic field ranging from $\sim(5-9)\times10^{12}$ G.
\end{abstract}


\section{Introduction} \label{sec:intro}
\par
Neutron-star X-ray binaries appear as the brightest objects in the X-ray sky. X-ray binaries can be classified as high-mass X-ray binaries (HMXBs) or low-mass X-ray binaries (LMXBs) according to the donor star masses. Based on the spectral type of the donor star, the neutron star HMXBs are also classified as either supergiant X-ray binaries or Be/X-ray binaries \citep{fornasini2023highmass}. A majority of the HMXB systems are known to be Be/X-ray binaries (BeXBs) in which young optical companions are spectral type O or B \citep{caballero2012x}.
An amount of emission in the infrared band is also observed from these companion stars. The observed emission lines and infrared excess are attributed to the presence of an equatorial disc around the Be star, 
which is formed by the rapid rotation of the Be star expelling material \citep{Porter_2003}. The outburst activity of 
transient BeXBs is usually divided into two types according to the X-ray variability in terms of duration and luminosity. 
Type I X-ray outbursts exhibit regular and periodic (or quasi-periodic) behaviour, occurring close to the periastron 
passage of the neutron star \citep{stella1986intermittent}. Type II X-ray bursts are major events that 
represent a significant increase of $10^3-10^4$ times of the X-ray flux compared to the quiescence and are normally known to 
last for several weeks to months \citep{refId0}. These outbursts are possibly caused by the enhanced episodic outflow of the 
Be Star \citep{paul2011transient}. 

The high-energy radiation associated 
with the rotation of the neutron star is produced as a result of the gradual accumulation of matter from a donor star heating the polar cap 
of the neutron star. At low luminosity, radiation is produced by hot spots or mounds on the polar cap and predominantly 
escapes along magnetic field lines, resulting in the so-called ``pencil beam'' \citep{burnard1991accretion,nelson1993nonthermal}. 
At high luminosity, radiation pressure is sufficient to effectively decelerate the accreted plasma,
forming a radiatively dominated shock above the polar cap and an extended emission region known as the accretion column. 
In this case, the radiation mainly escapes through the column walls, forming a “fan beam” \citep{basko1976limiting,wang1981analysis}. 

\par
The pulse profile of some sources exhibits a complex evolution with energy and luminosity, in particular around the expected
critical luminosity, which may be related to the change in emission patterns \citep{wang2022timing,ji2020switches}. 
Pulse profiles are also observed to change near the cyclotron frequency in some sources (e.g. V0332+53, 4U 0115+63, 
1A 0535+262) \citealt{lutovinov2009timing}). The pulse fraction, a measure of the pulsation amplitude depending on the energy band, has been analyzed in X-ray pulsars. The pulse fraction of some sources is known to exhibit a complex evolution 
with energy \citep{tsygankov2010completing}. The characteristics of the accreting plasma may undergo abrupt changes around
the cyclotron frequency, leading to alterations in the observable emitted beam pattern \citep{lutovinov2009timing}. 
The relationship between pulse fraction and energy, as dependent on luminosity, was observed in V0332+53 and 
4U 0115+63 \citep{tsygankov2010completing,tsygankov20074u}. The CRSF-dependent structure in the pulse fraction–energy 
relationship of 1A 0535+262 is the first source for which dependence is observed within a limited luminosity range 
between $4.8\times 10^{37}$ and $1.0\times10^{38}$\lum \citep{wang2022timing}.
\par 
Be/X-ray binary pulsar \exo was discovered by $EXOSAT$ observatory
during a giant X-ray outburst (type II outburst \citealt{1989ApJ...338..373P}). In this system, the pulsar with a spin period of 42 s \citep{fu2023timing} orbits a B0 Ve star \citep{janot1988optical} along with an
orbital modulation of 46 days \citep{Wilson_2005}. The system was located at a distance of 7.1 kpc 
measured by optical and infrared observations \citep{motch1987optical, 10.1093/mnras/232.4.865}, but the latest measured 
distance is $3.6_{-1.3}^{+0.9}$ kpc by Gaia \citep{10.1093/mnras/stab345}. Type I outbursts have been nearly detected at
every periastron passage of its approximately $\sim$46 days orbit period. \citep{wilson2008outbursts}. In June 2006, 
\exo was observed to experience the second giant outburst since its discovery with source flux peaking up to $\sim$750 mCrab from the \integral observatory \citep{klochkov2008giant}. In 2021, the third giant outburst with peak flux up to $\sim$550 mCrab has been monitored with the X-ray instruments onboard MAXI/GSC, NICER, Fermi/GBM, Swift/BAT, NuSTAR, \hxmt and IXPE observatories \citep{nakajima2021maxi,thalhammer2021nicer,tamang2022spectral,
fu2023timing,malacaria2023polarimetrically}. \cite{fu2023timing} showed the pulse profile evolution during the outburst based on \hxmt data. IXPE observations suggest a low polarization degree and the magnetic axis of \exo swings close to the observer’s line of sight \citep{malacaria2023polarimetrically}. 

\par
The study of pulsars' energy spectrum during outbursts provides a rather accurate method for the estimation of the 
physical properties of compact stars. Detection of cyclotron resonance scattering features (CRSFs) due to the photons being resonantly scattered by the electrons in the pulsar spectrum provides a direct way to estimate the surface magnetic field of neutron stars. The continuum spectrum in \exo was fitted by an acceptable continuum of the power law 
along with thermal blackbody component at 1.1 keV during the 1985 giant outburst \citep{sun1994x}. The majority of data of 
later observations were fitted by an absorbed power-law modified with a high energy cutoff model \citep{wilson2008outbursts}. 
\cite{refId0} find that NPEX also provided a statistically acceptable fit to the data. On using the continuum model \cpl, 
\cite{tamang2022spectral} also successfully fit the spectrum. Several authors discovered possible evidence for an absorption feature near 10 keV \citep{wilson2006detection,klochkov2007integral,wilson2008outbursts}. 
\cite{reig1999x} reports that a possible spectral absorption feature at 36 keV is tentatively attributed to a 
cyclotron absorption line. \cite{klochkov2008giant} suggested that the absorption line at $\sim$63\kev\ which appeared in the spectrum obtained close to 
the maximum of the 2006 outburst might be the first harmonic line. Thus, there has not solid evidence of CRSFs in \exo until now.
\par
In this paper, we report the detailed results of the timing and spectral analysis of the broadband spectrum (2--105\kev) of the
X-ray pulsar \exo during the 2021 giant outburst observed with the \hxmt. In Section \ref{OBS}, we present the observations 
and the data extraction. Timing analysis and pulse profiles are presented in section \ref{sec:Timing}. The X-ray spectral analysis, 
including phase-averaged spectrum and phase-resolved spectroscopy, are presented in section \ref{Spe}, and we report the discovery of the cyclotron resonance scattering 
feature in \exo. Broad continuum spectral fitting
with a Comptonization model is presented in section \ref{Comptonization}. The conclusion and a brief discussion are summarized in Section \ref{discuss}.

\section{OBSERVATIONS} \label{OBS}

\hxmt is the first X-ray astronomical satellite in China, launched on 2017 June 15. \hxmt consists of 
three main instruments: the High Energy X-ray telescope (HE) operating in 20--250~\kev\ and the areas of the telescopes are 5100 cm$^2$ \citep{liu2020high}, the Medium Energy X-ray telescope (ME) operating 
in 5--30\kev\ with a detection area of 952 cm$^{2}$ \citep{cao2020medium} and the Low Energy X-ray 
telescope (LE) covering the energy range 1--15\kev with a detection area of 384 cm$^2$ \citep{chen2020low}.

\begin{figure}
    \centering
    \includegraphics[width=.5\textwidth]{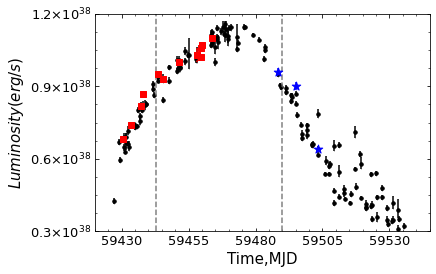}
    \caption{The evolution of luminosity (2–105 \kev) is estimated by fitting the \hxmt spectra. The red square points, blue star points represent where cyclotron absorption line around 47 keV
    may be evident in the rising and fading phases during the outburst respectively.}
    \label{fig:1}
\end{figure}

\begin{table*}
\centering
\caption{List of \hxmt observations of the pulsar \exo during 2021 Type-II outbursts}
\resizebox{0.85\textwidth}{!}{%
\begin{tabular}{lllll}
\hline
Observatory   & Year of Observations & Proposal ID & No. of Obs              & On Source time (\ksec)   \\ \hline
\hxmt         & 2021 Jul -- 2021 Nov  & P0304030    & 65                      & 2292                     \\ 
              &                      & P0404147    & 1                       & 172.5                    \\\hline
\end{tabular}%
}
\label{table1}
\end{table*}

\par
\exo was observed during the Type-II outburst from July 28, 2021 to November 21, 2021. We have used 66 observations
from \hxmt with a total exposure time of 2464.5 ksec. The specifications related to 
the \hxmt observations under consideration are presented in Table \ref{table1}. The Insight-HXMT Data Analysis Software 
(HXMTDAS) v2.04 is used to analyze data (more details on the analysis were introduced in previous publications, 
e.g., \citealt{WANG20211}; \citealt{chen2021relation}). In order to take advantage of the best-screened event file 
to generate the high-level products including the energy spectra, response file, light curves and background files, 
we use tasks $he/me/lepical$ to remove spike events caused by electronic systems and $he/me/legtigen$ be utilized to select 
good time interval (GTI) when the pointing offset angle $< 0.04^\circ$; the pointing direction above earth $> 10^\circ$; 
the geomagnetic cut-off rigidity $>8$ GeV and the South Atlantic Anomaly (SAA) did not occur within 300 seconds.
We also utilize the {\emph FTOOL} $grppha$ to improve the counting statistic of the spectrum. 

\par
In Figure \ref{fig:1}, the X-ray light curves of \exo which present the Type II outburst lasting about four months 
monitored by \hxmt are shown, and the pointing observations cover the entire outburst from 
half of the peak luminosity in ascending interval to about a quarter of the peak luminosity in the decrease phase. 
Thus, the good spectral analysis ability of the \hxmt and high-quality data during outburst allowed us to study 
the spectral variations of \exo and search for the possible cyclotron absorption features. 

\begin{figure}
    \centering
    \includegraphics[width=.48\textwidth,height=6.8cm]{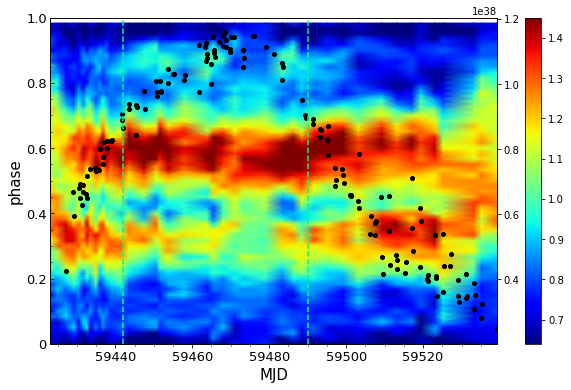}
    \caption{The evolution of the pulse profiles with time in the energy ranges 10.0--20.0\kev. 
    The color bar displays the values of the pulse profile, which are normalized by the pulse average count rate. 
    The green dotted lines at MJD 59442 and MJD 59490 indicate the epochs around the critical luminosity.}
    \label{fig:2}
\end{figure}

\section{TIMING ANALYSIS}\label{sec:Timing}
Based on high-cadence observations and high-quality data of \exo during its giant outburst in 2021, 
at first, we use HXMTDAS task $hxbary$ to change the photon arrival time from TT (Terrestrial Time) 
to TDB (Barycentric Dynamic Time) which considers the time delay due to the movement of the earth 
and satellite. The epoch-folding FTOOL $efsearch$ helps us to estimate the pulse period of $41.270\pm0.613$~sec for the example observation ObsID P030403002703 (MJD 59463). The range of errors is estimated by folding the light curve with a large number of periods around the approximate period by \chisq maximization. The uncertainties of the spin period are estimated using a Gaussian error. Using the obtained pulse period, we generated the light curves of each \hxmt observation by folding the 
background-subtracted light curves with a phase bin of 64.

The pulse profiles during the outburst evolution for the \hxmt ME detectors from 
10--20\kev\ are presented in Figure \ref{fig:2}. Pulse profiles show abrupt changes from double
peaks to a single peak around MJD 59442. The structure of peak 1 (0.35 phase) disappears and the intensity of 
peak 2 (0.6 phase) gradually improves. The pulse profile switches between single and double peaks around MJD 59490, 
corresponding to a luminosity of $0.8\times 10^{38}$\lum (assuming the distance of 7.1 kpc in this work). This luminosity also corresponds to the location of pulse 
profile transitions around MJD 59442. This phenomenon is consistent with investigations of this source \citep{fu2023timing}. 
The transition between the double peak and the single peak reveals the transition from the subcritical to supercritical 
accretion regime, corresponding luminosity can be interpreted in the context of the critical luminosity \citep{becker2012spectral}.

The pulse profiles of the accretion pulsar with energies are also studied. The entire \hxmt energy band was resolved into various sub-intervals as: 
2--5\kev, 5--7\kev, 7--10\kev, 10--15\kev, 15--20\kev, 20--25\kev, 25--30\kev, 30--35\kev, 
35--40\kev, 40--42\kev, 42--44\kev, 44--50\kev, 50--60\kev\ and 60--80\kev. Corresponding to 
each interval, the energy-resolved pulse profiles were generated in order to analyze the dependence 
of the shape with energy.
\par
\begin{figure}
    \centering
    \includegraphics[width=.48\textwidth,height=10cm]{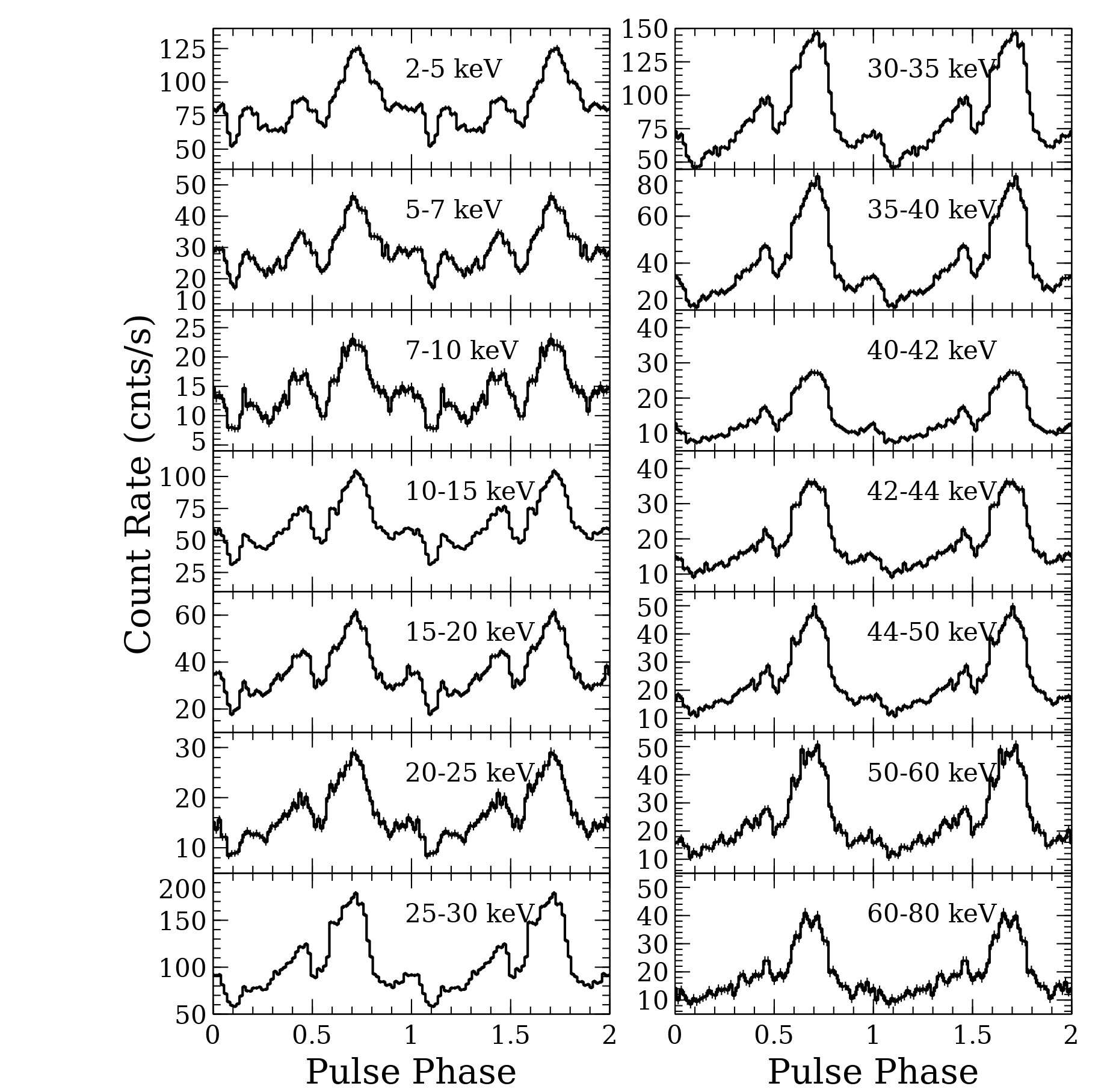}
    \caption{Pulse profile shapes as a function of the energy band for the 
    ObsID: P030403002703 (MJD 59463). Two pulse periods in each panel are presented.}
    \label{fig:3}
\end{figure}

The pulse profiles strongly depend on energy as the example presented in Fig. \ref{fig:3}. 
The shapes of the pulses below 20 keV appear to exhibit clearer indications of multi-peak patterns compared to those above 20\kev, which is observed for the majority of bright XRPs \citep{shaw2009accretion}. 
At the few to about 10 keV energy range, the pulse profile shows two distinct peaks: the main peak and the secondary peak at phases 0.4 and 0.7 respectively, and at about 0.2 phase and 0.95 phase,
there are two weak peaks in the pulse profile. A secondary peak, approximately $70\%$ of the intensity 
of the primary peak, is especially notable between 2--30\kev. Above 30 keV, the pulse profile still shows signs of double peaks with a secondary peak intensity decreasing with the energy, and above 60 keV, the pulse profile 
evolves from double peaks to a single peak. \cite{klochkov2008giant} analyzed the giant outburst data of \exo 
that was observed by JEM-X and IBIS/ISGRI in June–September 2006,  
and the evolution of the pulse profiles with energy is consistent with our result. 

\begin{figure}
    \centering
    \includegraphics[width=.4\textwidth,,height=10cm]{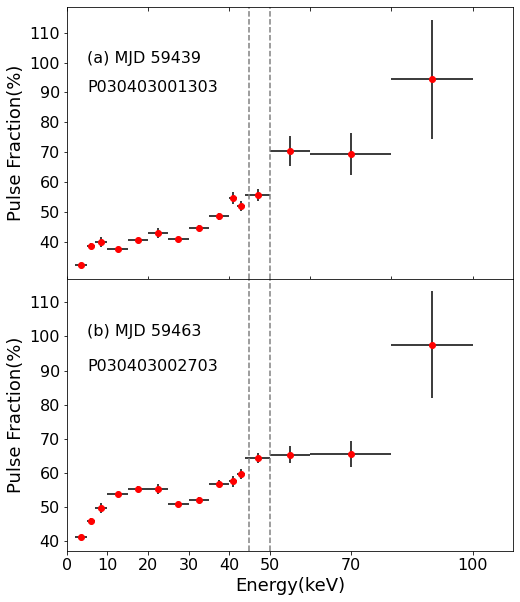}
    \caption{The pulse fraction of \exo as a function of energy for two observations: ObsID P030403001303 (MJD 59439); ObsID P030403002703 (MJD 59463)}
    \label{fig:4}
\end{figure}

We use the rms pulse fraction to characterize the observed changes in the pulse profile. The rms pulse fraction 
is determined as $\left( F_{\max}-F_{\min} \right) /\left( F_{\max}+F_{\min} \right)$, where $F_{\max}$ 
and $F_{\min}$ are fluxes in the maximum and minimum of the pulse profile, respectively. The error range is estimated by the propagation of uncertainties. We present the results of rms pulse fraction with energy
in Figure \ref{fig:4}. At MJD 59463 when the luminosity is high (around $1.1\times 10^{38}$\lum), 
the rms pulse fractions show a gradually rising trend below 20\kev. Between 20 and 30\kev, the rms pulse 
fraction shows a decreasing trend. Above 30\kev, the rms pulse fractions gradually rise as expected for 
accreting pulsars. As the luminosity decreases (e.g., at MJD 59439), the rms pulse fraction energy dependence again simplifies, 
with the rms pulse fraction rising gradually with energy between 1.0 and 100.0\kev, similar to that generally 
observed in accreting pulsar systems. Based on the assumption that high-energy photons 
are emitted from regions close to the neutron star surface, while soft photons are formed in the upper part of 
the column, the contrast between the minimum and maximum visible surfaces of the accretion columns is the highest 
at higher energies and exhibits a sharp decrease around cyclotron line energy \citealt{lutovinov2009timing}. 
The results obtained are also similar to those 
obtained by other authors (see, e.g., \citealt{klochkov2008giant}). 

\section{Spectral studies} \label{Spe}
\subsection{Phase-averaged spectrum} \label{avg-spectra}
We analyzed phase-averaged spectra of \exo by using data from \hxmt observations to probe spectral 
characteristics corresponding to change with luminosity and whether the cyclotron line features are presented in the pulsar spectrum. The spectral models that have been generally used to represent
the X-ray spectra of accreting X-ray pulsars are usually composed of a power-law continuum with low-energy 
absorption and a cut-off at energies about 12--20\kev. The spectrum of \exo is typical for accreting XRPs \citep{klochkov2008giant}. Several authors mentioned that a typical spectrum usually 
consists of a cutoff power-law component with cutoff energies around 20--30\kev, and the soft X-ray component 
could be fitted by a blackbody-like spectrum with temperature $kT\sim1.1$ \kev \citep{sun1994x}. 
\cite{naik2013timing} described the source spectrum by a two-component continuum model consisting of a blackbody component with temperature 1.1 keV and power law with an exponential cutoff.

Thus we successfully used power-law with a $\hcut$ model to fit the source spectra, 
despite this model leads to a line-like feature in the spectral around the $E_{c}$ \citep{Burderi_2000}. 
The function of the model is shown below:
\begin{equation}
f(E) = KE^{-\Gamma} \times 
\left\{
\begin{array}{ll}
1 & \text{if } E \leq E_c \\
\exp\left(-\frac{(E-E_{c})}{E_{f}}\right) & \text{if } E > E_{c}.
\end{array}
\right\} ,
\end{equation}
where $f\left( E \right)$ represents the high energy cutoff power-law model, $E$ is the photon energy and $K$ 
is the normalization factor, $\Gamma$ is the photon index of the power law, $E_{f}$ and $E_{C}$ is exponential 
folding energy and cutoff energy in units of keV respectively. The other form of the continuum model represents 
a power law with high energy exponential roll-off (\cpl\ model)
\begin{equation}
f\left( E \right) =KE^{-\alpha}\exp \left( -E/\beta \right) 
\end{equation}
where $\alpha$ is the power law photon index, E is the photon energy and K is the normalization factor. 
E-folding energy of exponential rolloff (in \kev) is described by $\beta$.

\par
We also modeled the data using a global absorption column $\tbabs$ model proposed by \cite{wilms2000} 
to describe the absorption of X-rays below $\sim$4\kev\ by gas and dust composed mainly of hydrogen 
in the Galaxy. There also exists an iron fluorescence emission line with an equivalent width of 200 eV 
at $\sim$6.4\kev, we add a Gaussian function to fit the iron emission line.

Based on the observations of \exo during the brighter portion of the 2006 June outburst, several authors had reported a cyclotron feature 
near 10\kev \citep{wilson2008outbursts,klochkov2007integral}, they added a gaussian absorption at $\sim$10\kev\ 
into a power law/cutoff model and the cyclotron energy was 10.1(2)\kev with a Gaussian width of 3.3(2)\kev and 
a peak depth of 1.1(1). Recently, \cite{tamang2022spectral} used the continuum model combination: 
\constant*\phabs*(\cpl+\gauss), and the spectrum showed highly significant negative residuals at about 10\kev,
the value of gabs strength and $\sigma _{gabs}$ is about 0.2\kev\ and 2\kev\ respectively. \cite{ferrigno2016glancing} mentioned that the source \exo shows some complex absorption features in its 
spectrum and cannot be modeled by a single continuum model. Therefore, we also modeled broadband continuum spectra of \exo with alternative spectral features 
to probe related negative residuals. 

\begin{figure}
    \centering
    \includegraphics[width=.5\textwidth]{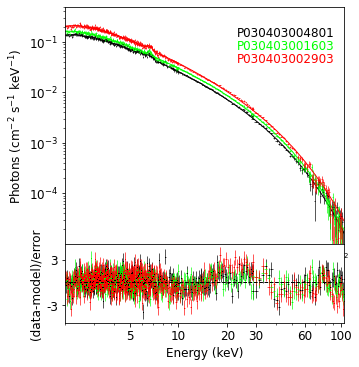}
    \caption{Phase-averaged energy spectra of \exo at different luminosity levels, obtained from 
    three epochs of HXMT observations during Type II X-ray outbursts. The spectra were fitted with 
    \cpl\ model multiplied by \tbabs~ along with an iron emission line at $\sim$6.4\kev. 
    The source spectrum and best-fit model are shown in the top panel whereas the contribution 
    of residuals to \chisq at each energy bin are shown in the bottom panel for each epoch of HXMT observations. The spectrum from ObsID P030403002903 shows an absorption feature around 10 keV.}
    \label{fig:5}
\end{figure}

\begin{figure}
    \centering
    \includegraphics[width=.47\textwidth]{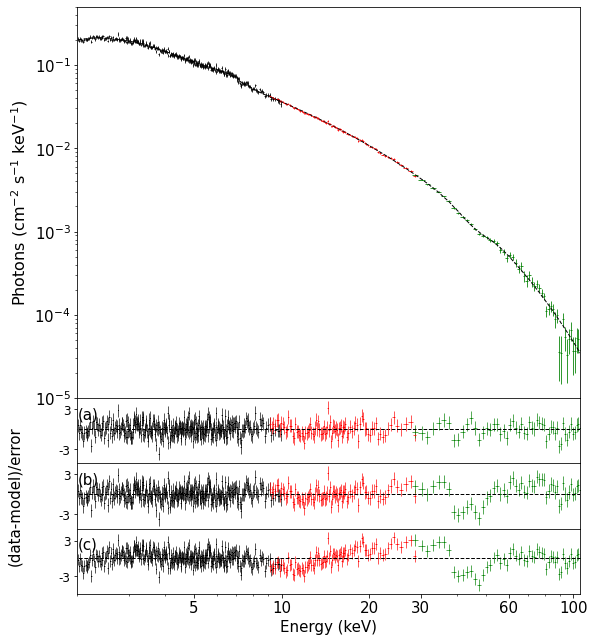} 
    \caption{Fitted spectra of \exo with the energy range from 2 to 105\kev\ obtained by \hxmt for ObsID: P030403002703). Black, red, and green data points represent the spectra from the three main instruments of Insight-HXMT corresponding to low energy (2-10 keV), medium energy (9-29 keV), and high energy (27-105 keV) respectively. The top panel presents the spectrum fitted with a continuum model \cpl\ plus a blackbody (resolve 10 keV feature) and gaussian $FeK\alpha$ at $\sim$6.4\kev\ 
    multiplied by \tbabs\ and a cyclotron scattering line at $\sim$47\kev\, and 
    corresponding spectral residuals (panel a) are shown. Panels b, c indicate the spectral 
    residuals obtained by fitting pulsar spectra with \cpl\ multiplied by \tbabs\ along with an iron emission line plus a 
    blackbody and without blackbody respectively.}
    \label{fig:6}
\end{figure}

\par
After fitting all the observations, we only explore the 10\kev\ absorption features in the spectra in seven observations around the peak of outburst with ObsIDs: P030403002401 (MJD:59454), P040414700104 (MJD:59459),
P030403002703 (MJD:59463), P030403002903 (MJD:59465), P030403003002 (MJD:59466), 
P030403003201 (MJD:59468), P030403003801 (MJD:59479). 
The energy spectra with the simple model consisting of: \tbabs*(\cpl+\gauss) are shown in Figure~\ref{fig:5}, which shows the energy spectra from of the three examples together, and black, red, and green colors correspond to an unabsorbed luminosity of $7.38\times 10^{37}$\lum, $1.10\times 10^{38}$\lum, and $8.64\times 10^{37}$\lum\ in the 2--105\kev\ range, respectively. Negative residuals can be observed at around 10\kev\ in the residual plot only for one ObsID P030403002903 represented in red, indicating a corresponding high luminosity of $1.10\times 10^{38}$\lum. Such features were 
not detected during the decaying and rising phases of the outburst. Various combinations of models 
such as: high energy cutoff power-law, cutoffpl, along with blackbody were used to test the reliability of the reported dip in seven observations. When we add a blackbody component with
$kT \sim$1 keV to the above continuum models, any signature of the broad absorption feature at $\sim$10\kev\ was not found in the residuals, and a reduced-\chisq 
close to 1 (see an example in Fig. \ref{fig:6}). 
For \exo, if we allowed for another free absorption line at near 10\kev, however, the fit `line' was a broad feature near 6.9\kev. Thus, a possible cyclotron absorption line would be model-independent and should 
appear in spite of change in continuum model used to fit the spectrum. A broad absorption-like 
feature at $\sim$10\kev\ was detected in some accreting pulsars (e.g., 4U 1907+09, Her X-1, \citealt{manikantan2023investigation}). \cite{epili2017} proposed that they used a high energy 
cutoff model along with a blackbody component to test the reliability of the reported line. The feature was also not detected in the pulsar spectra obtained from RXTE observations when the blackbody component was included
\citep{epili2017}. Therefore, we conclude that the pattern of 10 \kev~residuals is not a signature of the cyclotron resonance scattering feature obtained from all the observations 
during peak of outburst.
\par
For the details of the spectral fitting for the observations with possible 10 keV features, the example spectrum and fitting procedures are shown in Fig. \ref{fig:6}. The initially adopted model: ``\tbabs*( \cpl+ \gauss)'', does not give a good fit for ObsID P030403002703, specially the absorption feature at $\sim$47\kev, negative residuals can be also observed at $\sim$10\kev\ as relevant from the bottom panel of 
Fig. \ref{fig:6} and \chisq \( d.o.f \) value of 1441(1162). We then added a blackbody component to improve the fits, the 10 \kev\ residuals disappear, the \chisq changed to 1219 (1160 d.o.f.). The absorption 
structure around 40--55\kev\ still be observed in the middle residual panel of Fig. \ref{fig:6}. A possible explanation for those residuals at $\sim$47\kev\ is that there might be presence 
of cyclotron line which allowed us to add a CRSF component in the spectral model. In the case 
of the model: \tbabs*(\cpl*\gabs+\gauss+\bbody), the \chisq changed to 1129 (with 1157 d.o.f) with a F-test probability of $4\times 10^{-19}$ for the fitting improvement. The low false alarm probabilities may make the detection of the line stable against even crude mistakes in the computation of the significance \citep{kreykenbohm2004x}. Therefore, we conclude that a statistically acceptable spectral fit can be obtained when a Gaussian absorption line is included in the spectral model. For other observations without 10 keV features, the fitting results by adding the blackbody component did not improve, then we did not consider adding the blackbody component for these observations without a 10 keV absorption feature.

\par
\begin{figure}
\includegraphics[width=.48\textwidth]{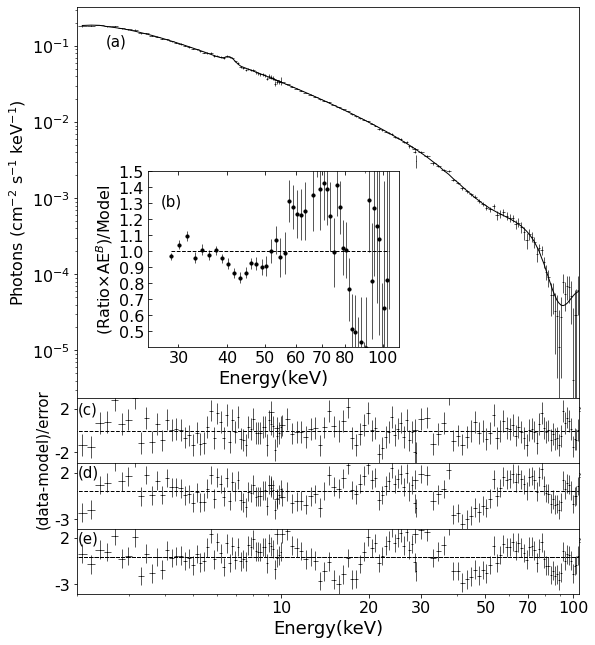}
\caption{The 2-105\kev energy spectrum of \exo obtained from the ObsID P030403002001. The spectra along with the best fitting model e.g. $\cpl$ multiplied by $\tbabs$ plus two absorption lines at $\sim45$ and $\sim90$ keV along with a Gaussian for iron emission line (panel a) and corresponding spectral residuals (panel c) are shown. Panels d, e indicate the spectral residuals obtained by 
fitting pulsar spectra with cutoffpl and high energy cutoff model, respectively, along with 
interstellar absorption component and a Gaussian function for iron emission line at 6.4 keV. Any signature of cyclotron absorption line at previously reported value of $\sim 10$ keV is not seen in the spectral residuals. The Crab-ratio (panel b) exhibits a noticeable negative 
feature from 40-55 keV. Although the ratio showed an absorbed line at 90 keV, the spectrum above 60 keV has much poorer statistics.}
\label{fig:7}
\end{figure}

As noted in Fig.\ref{fig:1}, the part of observations show the absorption feature around 47\kev. During the peak region, the spectrum of one observation on Aug 24, 2021 (ObsID P030403002001), shows two possible absorption features, thus here 
we show detailed spectral fits for this representative observation in Fig.  \ref{fig:7}. Following \cite{wilson2008outbursts}, we first use a model combinations: 
\tbabs*(\powerlaw*\hcut+\gauss)  to describe the energy spectrum of \exo. After the continuum spectral fittings, there exist the significant negative residuals at $\sim45-60$\kev\
and $\sim75-100$\kev\ (see Fig. \ref{fig:7}(e)). This model leads to a line-like feature in the spectrum around the $E_{c}$ \citep{Burderi_2000} and a poor fit with a reduced \chisq of 1.15. 
We used different model combinations: \tbabs*(\cpl+ \gauss) to fit the broadband spectrum of the source with the $\chi^2$ of 954 (861 d.o.f.), which still results in two absorption features: one absorption 
line at 45\kev\ and a weaker negative residual at 90\kev\ as observed from Fig.~\ref{fig:7}(d). We then added two additional absorption components (gabs) to improve the fits, the $\chi^2$ changed to 842 (855 d.o.f.). The spectrum of \exo along with the best-fitting model Fig.~\ref{fig:7}(a) and residuals Fig. \ref{fig:7}(c) for \hxmt observations are shown. 

\begin{table}
\centering
\caption{Best-fitting parameters of \exo observed with Insight-HXMT for ObsID P030403002001}
\begin{tabular}{lll}
\hline
Date                      & \multicolumn{1}{c}{}             & \multicolumn{1}{c}{2021-8-24} \\ \cline{1-1} \cline{3-3} 
MJD                       & \multicolumn{1}{c}{}             & \multicolumn{1}{c}{59450}     \\ \hline 
Model                     & \multicolumn{1}{c}{Parameters}   &                               \\ \hline \\
\tbabs                    & $N_{H}\,10^{22}\,\,cm^{-2}$      & $1.51^{+0.12}_{-0.05}$        \\ \\
\multirow{2}{*}{\cpl}     & $\varGamma$                      & $1.31^{+0.05}_{-0.02}$        \\
                          & $E_c\left( keV \right)$          & $31.20^{+5.81}_{-2.22}$         \\ \\
\multirow{3}{*}{\gabs}    & $\tau$                      & $0.36^{+0.20}_{-0.10}$          \\
                          & $E_{Gabs1}$(\kev)                & $48.2^{+1.8}_{-1.9}$          \\
                          & $\sigma_{Gabs1}$(\kev)           & $10.0^{+2.4}_{-1.9}$            \\ \\
\multirow{3}{*}{\gabs}    & $\tau$                      & $1.30^{+1.59}_{-0.37}$        \\
                          & $E_{Gabs2}$(\kev)                & $89.64^{+9.00}_{-2.28}$         \\
                          & $\sigma_{Gabs2}$(\kev)           & $8.15^{+6.62}_{-1.94}$          \\ \\
\multirow{2}{*}{\gauss}   & $E_{Fe}\left(keV \right)$        & $6.67^{+0.02}_{-0.1}$         \\
                          & $\sigma_{Fe}\left( keV \right)$  & $0.21^{+0.11}_{-0.05}$        \\ \\
Luminosity                & $L_{(2-105)keV}$($10^{38}$\lum) & $1.01^{+0.02}_{-0.01}$          \\  \hline
Fitting                   & $\chi_{red}^{2}(d.o.f)$          & $0.98(855)$                   \\ \hline
\end{tabular}
\label{table2}
\end{table}

\begin{figure*}
    \centering
    \includegraphics[width=.94\textwidth,height=5.4cm]{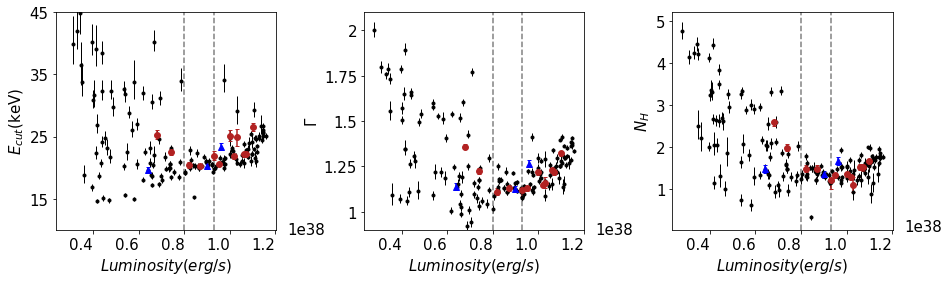}
    \caption{Luminosity evolution of spectral parameters such as cutoff energy (left panel), photon index (middle panel), column density (right panel), 
    obtained from the spectral fitting of HXMT observations of \exo 
    with a $\cpl$  multiplied by $\tbabs$ model during 2021 Type II outburst. The red circles and blue triangles indicate the presence of the cyclotron absorption line around 47 keV during the rising and fading phases of the outburst respectively}
    \label{fig:8}
\end{figure*}

\par 
In order to check the presence of two absorption-like features in the pulsar spectrum. We examined the ratio 
between the ObsID P030403002001 (MJD:59450) and Crab pulsar count rate spectra derived by HE detectors 
(see Fig. \ref{fig:7}(b)). This ratio has the advantage of removing the presence of any uncertainties related 
to calibration. We used the data of Crab observed by the \hxmt in September 2021 (MJD:59462), which is closest 
to the observation time of the ObsID P030403002001. This ratio exhibits a noticeable negative feature below 60\kev. 
Although the ratio showed an absorbed line at 90\kev, the spectrum above 60\kev\ has much poorer statistics, so the presence of this feature is unclear. 
\par
Therefore, the final model for the spectral fittings of all HXMT observations in \exo from $2-105$ keV can be summarized as
\[
I(E) = \tbabs * (\cpl + \gauss ) * \gabs.
\]
The best-fitted parameters obtained from spectral fitting based on Obs ID P030403002001 with optimal models (possible two absorption feartures) are 
presented in Table \ref{table2}. In Table \ref{table3}, the best-fitting parameters of all the other observations containing only the 47\kev\ absorption structure are presented together.

Spectral parameters such as photon index ($\Gamma$), cutoff energy, additional column density ($N_{H}$), obtained 
from a continuum and $\tbabs$ model fitting of all \hxmt observations of \exo with corresponding luminosity are shown in Figure \ref{fig:8}. All these parameters showed intriguing trends with luminosity, which has been explored earlier 
by \cite{epili2017}. In the figure, we can notice that the values of power-law photon index distribute in three 
distinct regions such as negative, constant, and positive correlations with source luminosity, suggesting a direct measure of spectral transition in \exo. At lower luminosity, the pulsar spectrum was relatively soft. A negative 
correlation between the powerlaw photon index and luminosity can be clearly seen for this regime. The value of the 
photon index was found to vary between 1.2 and 2. When the luminosity was in the range of $0.9\times 10^{38}$\lum, 
the distribution of values of the photon index did not show any dependence on source luminosity. With an increase of 
source luminosity, the photon index showed a positive correlation with luminosity. In our spectral fitting, the column density ($N_{H}$) 
and cutoff energy ($E_{cut}$) also show similar changes with luminosity.

\begin{figure}
    \includegraphics[width=.48\textwidth,height=11.5cm]{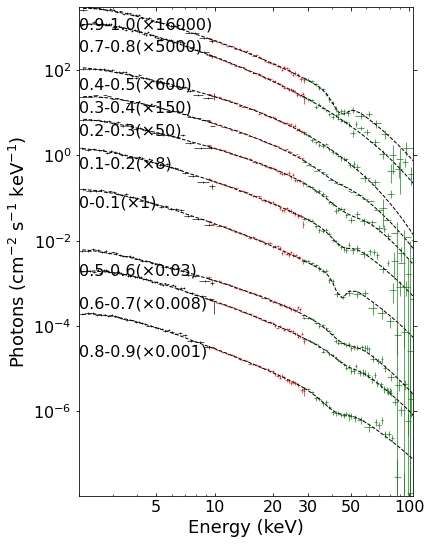}
    \caption{Phase-resolved spectral in different spin phases, obtained with the model: $\tbabs*\cpl*\gabs$ 
    for the ObsID P030403002001. The spectra of low-energy instruments  (2-10 keV), medium-energy instruments (9-29 keV), and high-energy instruments (27-105 keV), correspond to black, red, and green data points, respectively. A cyclotron absorption lines at $\sim 45$ keV is applied in the spectral fitting.}
    
    \label{fig:9}
\end{figure}

\begin{figure*}[htbp]
    \centering
    \includegraphics[width=0.45\textwidth,height=10cm]{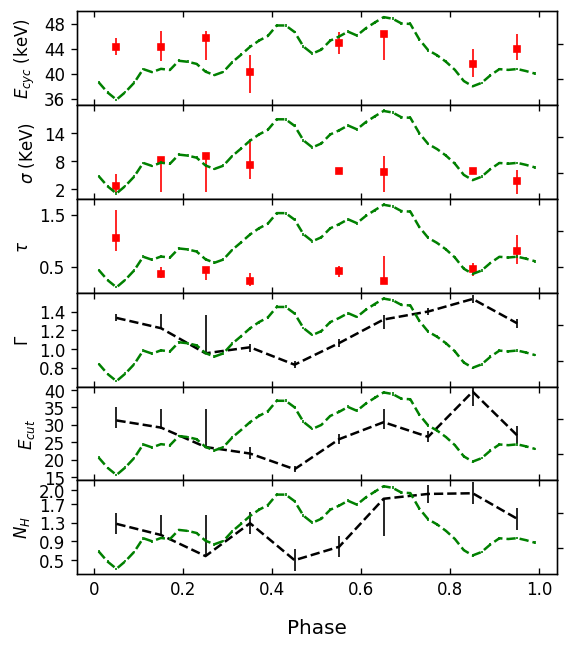}\hspace{0.4cm}
    \includegraphics[width=0.45\textwidth,height=10cm]{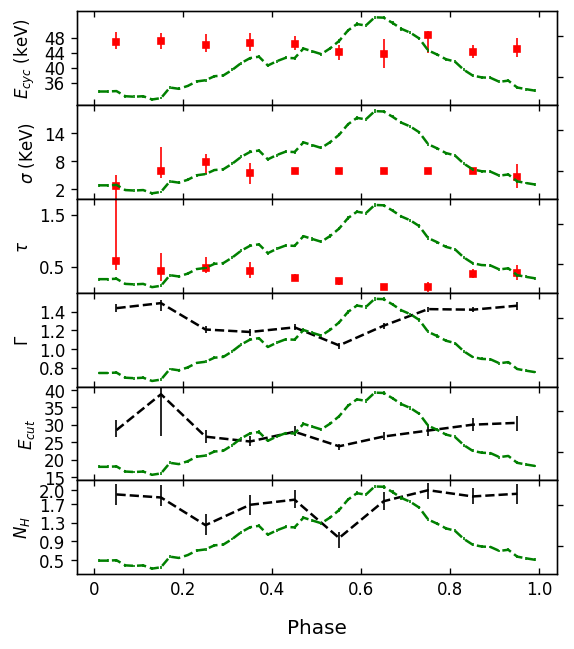}
    \caption{Pulse-phase variations in different spin phases, phase-resolved spectral fitting parameters of the cyclotron absorption lines and the continuum for 10 phase intervals, obtained with the cutoffpl and GABS models for two ObsIDs P030403002001 (Left panel) and P030403002703 (Right panel). The neutral hydrogen column density $N_{H}$ is in units of $10^{22}$ atoms $cm^{-2}$, $\Gamma$ is the photon index in cutoffpl. The centroid energy, width, of the cyclotron absorption lines are represented as $E_{cyc}$, $\sigma$, respectively, and the green lines in the panels indicate the pulse profile of ME (8-30 keV).}
    \label{fig:10}
\end{figure*}

\subsection{Pulse phase-resolved spectroscopy}\label{Spe-re}

This is well established that the spectrum of an X-ray pulsar varies with the pulse phase, parameters of cyclotron line 
and continuum would change with pulse phases. In order to reveal the phase dependence of the spectral features, we divide the pulse period into ten pulse phases, determine the spectrum for each phase, thus fit the phase-resolved spectrum with the model \tbabs*(\cpl)*\gabs. An example of such variability for 
the ObsID P030403002001 with individual pulse phase is presented in Fig. \ref{fig:9}. Variations in the best-fit parameter
values of the cyclotron absorption line over the pulse phase including two ObsIDs P030403002001 and P030403002703 are also presented in Fig. \ref {fig:10}. Here we focus on understanding whether the absorption features are detected at individual pulse phases in order to exclude the situation when 
the detection of the features in the averaged spectrum arises from the modeling of superimposed spectra variable across 
different pulse phases.
\par
The spectrum of the several phase intervals show obvious absorption features between 40\kev\ and 50\kev.  The inclusion 
of an absorption line at $\sim$47\kev\ improves the fit quality. After removing the gabs model from the overall model for ObsID P030403002001, 
the value of the \chisq changes from 731(742 d.o.f.) to 775(745 d.o.f.) at the $\varPhi _{phase}$ of 0.9-1.0. 
For phases 0.2-0.3, from 691 (674 d.o.f.) to 731 (677  d.o.f.). In the phases 0.2-0.3 and 0.9-1.0, the inclusion of the gabs model yielded the F-test probability were $2.9 \times 10^{-8}$ and $2.0 \times 10^{-9}$ respectively. 
In Figure \ref{fig:10}, the red points denote the 
line energy $E_{cyc}$, line width $\sigma$ and optical depth $\tau$ in the gabs model. For phases 0.5-0.6 and 0.8-0.9 of ObsID P030403002001, we fixed the width of the cyclotron line to be $\sim 6$\kev\ obtained from the average value of the other phases. Similarly, within the phases 0.4-0.9 for P030403002703, we maintained the width at 6 keV,
which may be due to the low S/N caused by the insufficient statistics of the phase-resolved spectrum. The lack of a 
significant absorption feature for phases 0.4-0.5 and 0.7-0.8 of ObsID P030403002001 does not motivate us to add a CRSF. We also find that $E_{cyc}$
varies between $40.3_{-3.4}^{+2.7}$\kev\ to $46.4_{-4.2}^{+0.4}$\kev\ along the pulse phase in 0.4–0.7. From Figure~\ref{fig:10}, 
the $\varGamma$, $E_{cut}$ and $N_{H}$ also show significant phase-dependent structures. $\varGamma$ exhibits a distinct 
bimodal pattern, being smaller than 1.0 during the main pulse and exceeding 1.2 elsewhere. This suggests a harder spectrum 
during the main pulse. $E_{cut}$ decreases from $\sim$35\kev\ to $\sim$20\kev\ when the phase changes from 0 to 0.5 and 
increases from $\sim$20\kev\ to $\sim$30\kev\ for phase 0.5-1.0. The neutral hydrogen column density components are also phase-dependent,
the hydrogen column density decreases by $0.7\times 10^{22}\,\rm atoms\,\,cm^{-2}$ within the phase 0-0.2, and it increases once again in the decay of the main peak.

\begin{table*}
\centering
\caption{Spectral fitting parameters for all observations containing cyclotron absorption lines}
\resizebox{\textwidth}{!}{
\begin{tabular}{llllllllllll}
\hline \\
ObsID & $N_{H}$($10^{22}$\pcm) & $bbody$(\kev) & \multicolumn{1}{c}{$\varGamma$} & $E_{c}$(\kev) &  \multicolumn{1}{c}{$\tau$}  & $E_{cyc}$(\kev) & $W_{cyc}$(\kev) & $E_{Fe}$(\kev) & $\sigma_{Fe}$(\kev) & $Luminosity$(\lum) & \chisq(d.o.f.) \\
 &  &  &  &  &  &  &  &  &  &  &  \\ \hline \\
P030403000501 & $2.41^{+0.12}_{-0.05}$ & - & $1.36^{+0.02}_{-0.01}$ & $27.4^{+0.6}_{-1.3}$ & $0.21^{+0.11}_{-0.08}$ & $43.8^{+4.3}_{-1.7}$ & $5.3^{+1.4}_{-2.8}$ & $6.4^{+0.10}_{-0}$ & $0.60^{+0}_{-0.09}$ & $6.79^{+0.13}_{-0.12}\times 10^{37}$ & $0.94(1330)$ \\ 
 &  &  &  &  &  &  &  &  &  &  &  \\ 
P030403000801 & $1.86^{+0.10}_{-0.08}$ & - & $1.23^{+0.02}_{-0.02}$ & $24.3^{+0.9}_{-0.9}$ & $0.16^{+0.08}_{-0.06}$ & $46.9^{+2.6}_{-3.2}$ & $8.4^{+3.0}_{-2.3}$ & $6.4^{+0.08}_{-0.09}$ & $0.56^{+0.02}_{-0.11}$ & $7.40^{+0.10}_{-0.10}\times 10^{37}$ & $0.84(1330)$ \\ 
 &  &  &  &  &  &  &  &  &  &  &  \\ 
P030403001103 & $1.50^{+0.09}_{-0.10}$ & - & $1.14^{+0.02}_{-0.02}$ & $22.4^{+0.9}_{-0.8}$ & $0.20^{+0.09}_{-0.08}$ & $47.5^{+2.1}_{-3}$ & $6.7^{+2.9}_{-2.9}$ & $6.6^{+0.02}_{-0.08}$ & $0.19^{+0.12}_{-0.06}$ & $8.22^{+0.12}_{-0.12}\times 10^{37}$ & $0.88(1330)$ \\
 &  &  &  &  &  &  &  &  &  &  &  \\ 
P030403001203 & $1.53^{+0.09}_{-0.09}$ & - & $1.16^{+0.02}_{-0.02}$ & $22.1^{+0.7}_{-0.7}$ & $0.18^{+0.07}_{-0.09}$ & $48.8^{+1.0}_{-2.8}$ & $8.3^{+2.6}_{-2.6}$ & $6.7^{+0}_{-0.15}$ & $0.13^{+0.11}_{-0.02}$ & $8.48^{+0.12}_{-0.12}\times 10^{37}$ & $0.98(1330)$ \\ 
 &  &  &  &  &  &  &  &  &  &  &  \\ 
P030403001702 & $1.35^{+0.08}_{-0.07}$ & - & $1.16^{+0.02}_{-0.02}$ & $22.2^{+0.8}_{-0.6}$ & $0.16^{+0.11}_{-0.06}$ & $49.1^{+4.8}_{-2.8}$ & $8.5^{+3.8}_{-2.0}$ & $6.6^{+0.09}_{-0.14}$ & $0.26^{+0.22}_{-0.12}$ & $9.22^{+0.10}_{-0.11}\times 10^{37}$ & $0.98(1330)$ \\ 
 &  &  &  &  &  &  &  &  &  &  &  \\ 
P030403001803 & $1.27^{+0.02}_{-0.21}$ & - & $1.13^{+0.01}_{-0.04}$ & $22.5^{+2.3}_{-1.7}$ & $0.19^{+0.10}_{-0.01}$ & $48.2^{+2.3}_{-1.7}$ & $11.6^{+3.9}_{-0}$ & $6.6^{+0.12}_{-0.12}$ & $0.22^{+0.12}_{-0.12}$ & $9.28^{+0.12}_{-0.13}\times 10^{37}$ & $1.04(1330)$ \\ 
 &  &  &  &  &  &  &  &  &  &  &  \\ 
P030403002101 & $1.35^{+0.06}_{-0.06}$ & - & $1.22^{+0.02}_{-0.02}$ & $25.1^{+0.9}_{-0.8}$ & $0.23^{+0.13}_{-0.07}$ & $48.9^{+3.6}_{-1.9}$ & $7.3^{+2.8}_{-2.2}$ & $6.6^{+0.02}_{-0.06}$ & $0.24^{+0.10}_{-0.06}$ & $1.01^{+0.01}_{-0.01}\times 10^{38}$ & $0.91(1330)$ \\ 
 &  &  &  &  &  &  &  &  &  &  &  \\ 
P030403002503 & $1.09^{+0.18}_{-0.16}$ & - & $1.16^{+0.02}_{-0.03}$ & $24.9^{+1.3}_{-1.5}$ & $0.35^{+0.12}_{-0.13}$ & $45.9^{+3.2}_{-2.5}$ & $6.7^{+2.0}_{-2.2}$ & $6.5^{+0.14}_{-0.2}$ & $0.31^{+0.23}_{-0.17}$ & $1.02^{+0.03}_{-0.03}\times 10^{38}$ & $1.07(1330)$ \\ 
 &  &  &  &  &  &  &  &  &  &  &  \\ 
P040414700103 & $1.31^{+0.08}_{-0.08}$ & - & $1.18^{+0.02}_{-0.01}$ & $23.6^{+0.5}_{-0.6}$ & $0.22^{+0.08}_{-0.04}$ & $46.51^{+2.0}_{-0.9}$ & $4.8^{+1.6}_{-0.5}$ & $6.7^{+0.01}_{-0.09}$ & $0.11^{+0.13}_{-0}$ & $1.02^{+0.01}_{-0.01}\times 10^{38}$ & $0.97(1330)$ \\ 
 &  &  &  &  &  &  &  &  &  &  &  \\ 
P040414700104 & $1.11^{+0.17}_{-0.14}$ & $1.19^{+0.04}_{-0.08}$ & $0.96^{+0.04}_{-0.03}$ & $19.5^{+0.6}_{-0.4}$ & $0.19^{+0.10}_{-0.06}$ & $46.9^{+3.7}_{-2.6}$ & $10.2^{+3.3}_{-1.7}$ & $6.6^{+0.08}_{-0.11}$ & $0.30^{+0.24}_{-0.09}$ & $1.06^{+0.01}_{-0.01}\times 10^{38}$ & $0.96(856)$ \\ 
 &  &  &  &  &  &  &  &  &  &  &  \\ 
P040414700105 & $1.50^{+0.07}_{-0.08}$ & - & $1.20^{+0.01}_{-0.01}$ & $24.1^{+0.5}_{-0.6}$ & $0.20^{+0.05}_{-0.06}$ & $48.9^{+0.9}_{-2.8}$ & $7.0^{+1.5}_{-2.0}$ & $6.57^{+0.06}_{-0.11}$ & $0.21^{+0.22}_{-0.07}$ & $1.07^{+0.01}_{-0.01}\times 10^{38}$ & $1.0(1330)$ \\ 
 &  &  &  &  &  &  &  &  &  &  &  \\ 
P030403002703 & $3.01^{+0.94}_{-0.71}$ & $0.78^{+0.07}_{-0.04}$ & $1.15^{+0.04}_{-0.04}$ & $22.3^{+0.8}_{-0.7}$ & $0.18^{+0.06}_{-0.05}$ & $45.5^{+1.8}_{-1.5}$ & $5.1^{+1.9}_{-1.7}$ & $6.42^{+0.13}_{-0.13}$ & $0.81^{+0.22}_{-0.19}$ & $1.13^{+0.01}_{-0.01}\times 10^{38}$ & $0.97(1157)$ \\ 
 &  &  &  &  &  &  &  &  &  &  &  \\ 
P030403004202 & $1.80^{+0.10}_{-0.09}$ & - & $1.29^{+0.02}_{-0.02}$ & $25.0^{+0.9}_{-1.0}$ & $0.20^{+0.12}_{-0.08}$ & $47.2^{+4.9}_{-2.9}$ & $7.4^{+3}_{-2.9}$ & \multicolumn{1}{c}{—} & \multicolumn{1}{c}{—} & $9.52^{+0.16}_{-0.10}\times 10^{37}$ & $1.06(979)$ \\ 
 &  &  &  &  &  &  &  &  &  &  &  \\ 
P030403004602 & $1.32^{+0.09}_{-0.09}$ & - & $1.15^{+0.02}_{-0.02}$ & $22.2^{+0.8}_{-0.7}$ & $0.18^{+0.09}_{-0.06}$ & $45.7^{+2.7}_{-1.9}$ & $8.1^{+2.7}_{-2.2}$ & $6.4^{+0.18}_{-0.01}$ & $0.47^{+0.10}_{-0.11}$ & $8.70^{+0.16}_{-0.10}\times10^{37}$ & $0.90(1330)$ \\ 
 &  &  &  &  &  &  &  &  &  &  &  \\ 
P030403005001 & $1.40^{+0.18}_{-0.04}$ & - & $1.18^{+0.04}_{-0.01}$ & $23.0^{+1.2}_{-0.1}$ & $0.28^{+0.12}_{-0.02}$ & $47.39^{+5.14}_{-1.1}$ & $11.3^{+3.5}_{-1.0}$ & $6.5^{+0.14}_{-0.07}$ & $0.63^{+0.06}_{-0.30}$ & $6.41^{+0.10}_{-0.10}\times10^{37}$ & $0.99(1330)$ \\ \\ \hline
\end{tabular}}
\label{table3}
\end{table*}

\section{Broad continuum spectral fit with a Comptonization Model} \label{Comptonization}

In order to explore the accretion column emission from the pulsar in \exo, during its 2021 Type-II outburst,
we have applied the Thermal \& Bulk Comptonization model (BW model) proposed by \citet{becker2007thermal} to study the broad X-ray spectra from 2 -- 105 keV. This model assumes that, the emergent radiation spectrum from the NS accretion column emission is a sum of bulk and thermal 
Comptonization of seed photons produced via bremmstrahlung, cyclotron and blackbody emission in the accretion plasma. 
The formation of blackbody seed photons is at the base of the column where the thermal mound is located. Whereas the
seed photons of cyclotron and Bremmstrahlung emission are produced above the thermal mound in optically thin regions.

The BW model is implemented in \xspec as \bwcycl\ whose application details are explained in \citet{Ferrigno2009}. 
Assuming the canonical NS mass and radii as $M_{NS}=1.4~M_{\odot}$ and $R_{NS}=10~R_{\odot}$, the BW model has left with six 
fundamental free parameters. These are namely: the mass accretion rate ($\dot M$), the electron temperature ($T_e$), 
the radius of accretion column ($r_{0}$), the magnetic field strength ($B$), the Comptonization parameter ($\delta$) and 
the photon diffusion parameter($\xi$). The parameters $\xi$ and $\delta$ are defined as:
\begin{equation}
\xi=\frac{\pi r_0 m_p c}{\dot M \left(\sigma_\parallel
\sigma_\perp\right)^{1/2}} \ , \ \ \ \ 
\frac{\delta}{4}=\frac{y_\mathrm{bulk}}{y_\mathrm{thermal}}, \ 
\label{eq:BW-xi-delta}
\end{equation}
where $m_p$ denotes the proton mass, $c$ is the speed of light,
and $y_\mathrm{bulk}$ and $y_\mathrm{thermal}$ represent the
Compton $y$-parameters for the bulk and thermal Comptonization
processes describing the average fractional energy change experienced by a photon before it escapes
through the column walls.

The X-ray luminosity is obtained from the observed source flux of \exo in the range of 1-100\kev. The source flux is 
estimated at first from the best fitted empirical high-energy cutoff power law model to the phase averaged spectra of 
\exo for each of the observation epochs as mentioned in Table \ref{table4}. We consider the distance to the source as 
7.1 kpc and assume an isotropic accretion with unitary efficiency (i.e observed X-ray luminosity being equal to the 
accretion luminosity). The mass accretion rate, $\dot M$  therefore can be constrained from the X-ray luminosity 
estimates as : $L{x}=\frac{G~M_{NS}\dot M}{R_{NS}}$. Its known that, $\dot M$ is  strongly degenerate with $r_{0}$ 
and the parameter $\xi$. Therefore, as suggested in \citet{Ferrigno2009}, we fix the value of $r_{0}$ after 
obtaining a good fit. A similar procedure has been carried out earlier by \citet{epili2017}. Except the model 
normalization, the other normalization components of the model due to blackbody, cyclotron and the Bremsstrahlung 
emission seed photon are kept fixed to the values as suggested for \bwcycl\ model in \xspec. For the spectra obtained 
at the the peak of the outbursts (i.e observations between MJDs 59466--59486), a partial covering component (\tbpcf) 
is needed along with the galactic absorption component (\tbabs) to explain the heavily absorbed spectra. We have 
noticed from the spectral residual that, there is an additional emission component near $6.4-6.6$\kev\ seen in the 
spectra due to emission for neutral Fe~$K\alpha$. This has been modelled with a Gaussian component as has been 
done with while fitting the phase averaged spectra with the empirical models in section~\ref{avg-spectra}. 

\begin{figure}
    \centering
    \includegraphics[width=0.44\textwidth,height=14cm]{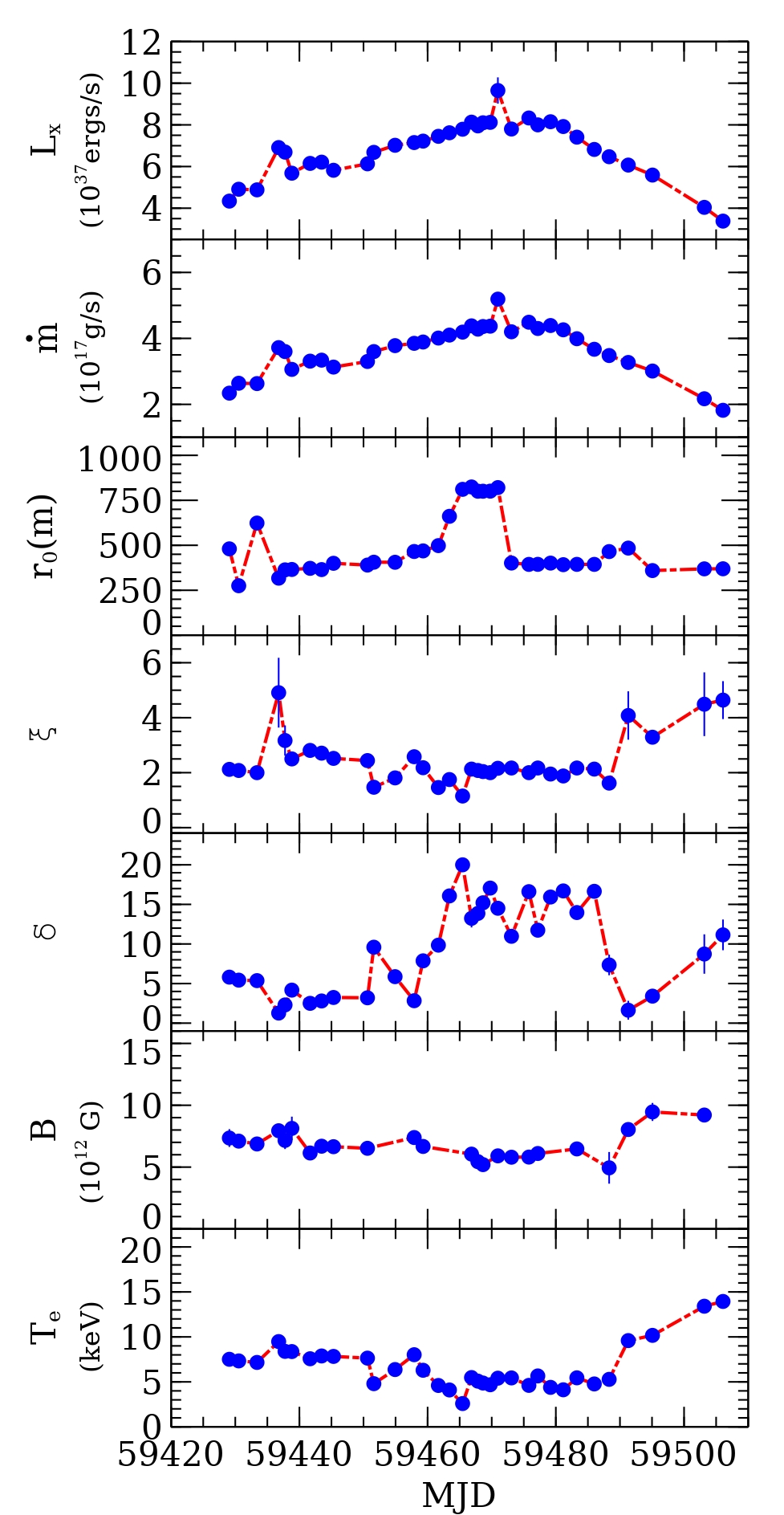}
    \caption{BW-model spectral parameter variations during the 2021 giant outburst of \exo with 
    varying source luminosity. A log of the best-fitted spectral parameters with the BW model is also shown in Table~\ref{table4}.}
    \label{fig:11} 
\end{figure}

\begin{figure*}
    \includegraphics[height=6 cm,angle=-90]{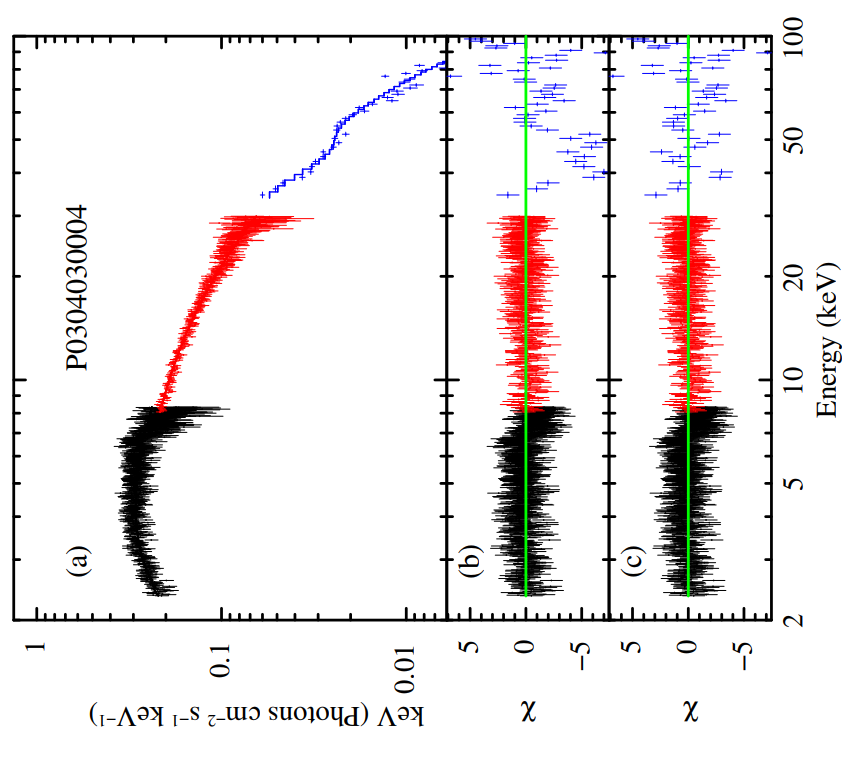}
    \includegraphics[height=6 cm,angle=-90]{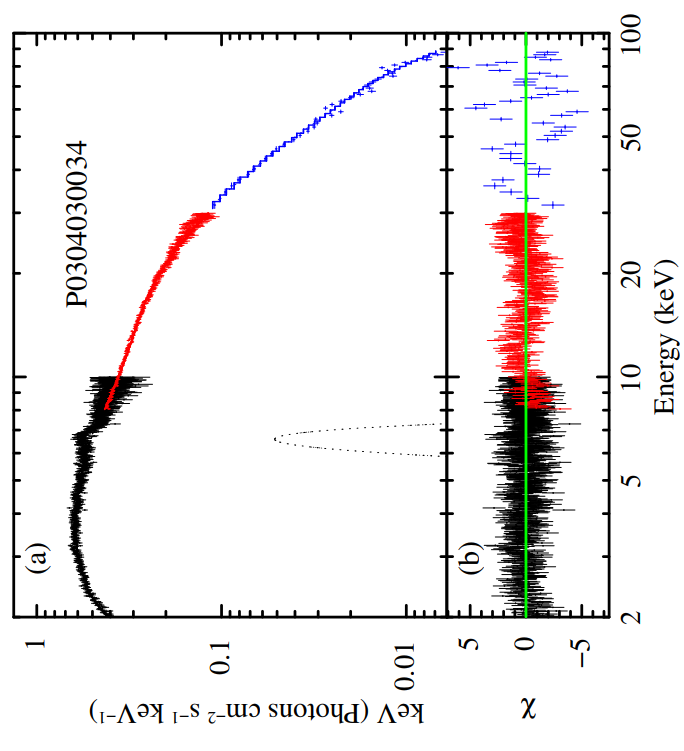} 
    \includegraphics[height=6 cm,angle=-90]{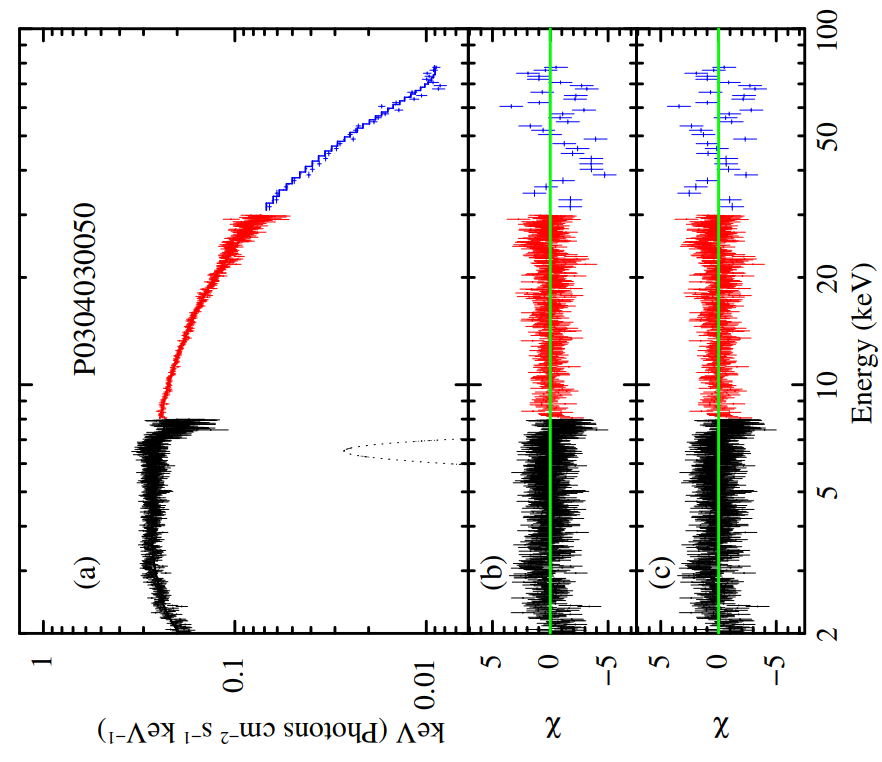} \\
    \caption{(a): The unfolded spectra of \exo in 2-105\kev obtained from three example observations during the 2021
    giant outburst. The spectrum in the left is obtained from ObsID: P0304030004 (near MJD 59429.08) before 
    the outburst peak. The spectrum shown in the middle is obtained from the ObsID:P0304030034 taken at the peak of the outburst (MJD: 59470.96). The spectrum shown in the right side is obtained from the declining phase of the giant outburst at MJD 59503.16 (ObsID:P0304030050).
    The spectral parameters obtained with the BW model for these observations are shown in Table~\ref{table4}.
    The spectral residuals shown in the panels (b) in each figure are obtained from the BW model without any
    cyclotron line component. Whereas the residuals shown in panels (c) are for best-fitted BW model requiring
    a cyclotron line component in hard X-rays . A Gaussian component near 6.4\kev is added to the spectra obtained at the peak and after the outburst peak to account for the emission from neutron Fe~$K\alpha$ 
    line near to the pulsar in \exo.}
    \label{fig:13}
\end{figure*}

In hard X-rays, we have used an absorption component (\gabs) between 40-55\kev\ to account for the cyclotron line absorption
seen in the spectra of \exo from the \hxmt observations. We have used the composite BW model along with the additional 
components to explain the spectra \exo at 34 luminosity epochs across its 2021 outburst observed with \hxmt. The best fitted
spectral parameters obtained BW model are shown in Table~\ref{table4}. It can be seen from the reduced-\chisq values that,
we could obtain a better fit of the phase-averaged spectra in wide luminosity range of \exo. In Figure~\ref{fig:11}, 
we show a variation of these physical parameters as the source luminosity varies across the each observation during the 
giant outburst. The best-fitted phase averaged spectra with BW model at three different luminosity epochs are shown 
in Fig.~\ref{fig:13}.

\begin{figure}
    \includegraphics[width=0.55\textwidth, height=9cm]{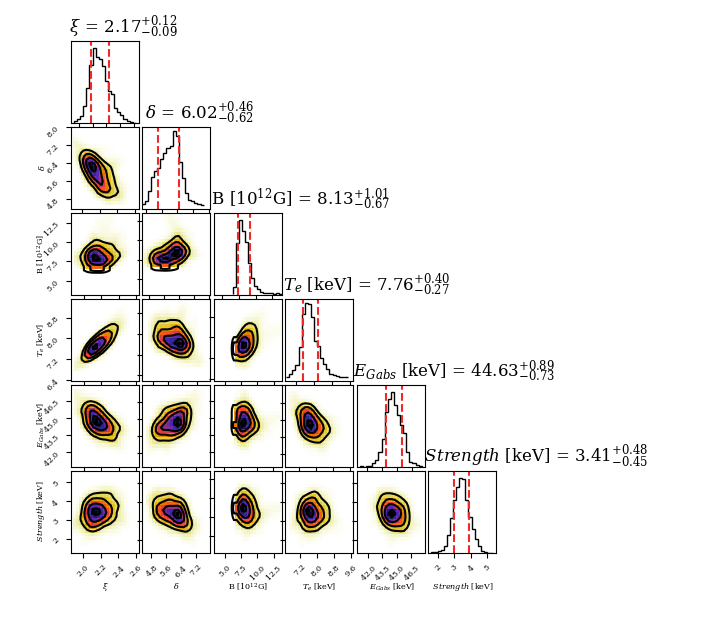}
    \caption{Corner plots obtained for the BW model bestfit spectral parameters of the ObsID:P0304030004. We have run MCMC chain with a 20000 samples (with first 10000 
    samples as burn length) of the best-fitted BW model along with a Gaussian absorption 
    component for the cyclotron line seen at $\sim44.7$\kev in the phase-averaged spectra.}
    \label{fig:12} 
\end{figure}

To check the parameter degeneracy of the best fitted BW model spectral parameters, we have run a Monte Carlo Markov Chain
(MCMC) having chain length of 20000  with Goodman-Weare chain algorithm using 20 walkers for the Obs-ID:P0304030004. 
A corner plot \citep{corner} obtained from  these \xspec MCMC chains is shown in Figure~\ref{fig:12}. During the 2006 
giant outburst of \exo, the broadband spectra in 3-60\kev were explained with a BW model \citep{epili2017} 
without the need of any cyclotron line component. However the present 2021 outburst observations clearly detected 
a strong cyclotron line component in broad hard X-rays from 2- 105 keV.

Among the BW spectral parameters we find that, the estimated range of NS magnetic field responsible for the production of 
cyclotron emission as per BW model is $(\sim 5-9)\times10^{12}$~G.  However, during the progress of outburst, we observed 
the varying of cyclotron line energy in the range of 40--55\kev. An estimated value of NS magnetic field from the range 
of detected cyclotron line energy is within $(\sim4.1-5.7)\times10^{12}$~G. This clearly indicates that the Comptonized 
emission region as per BW model is spatially offset from the cyclotron absorption region in \exo . Such an offset is also 
seen in the case of 4U~0115+63 \citep{Ferrigno2009}. We discuss more on this in Section~\ref{discuss}. During the onset, 
progress and peak of the giant outburst, the NS magnetic field is found to be within the range of $(5-8)\times10^{12}$~G.
However during the declining phase of the outburst as the source luminosity decreases due to decrease in mass accretion rate, 
we notice higher estimates of the magnetic field value increasing upto $\sim9.5\times10^{12}$~G. This indicates a luminosity 
dependence of the Comptonized  emission region in \exo. In other words, a change of column emission height with luminosity, 
from where the Comptonized emission of X-ray photons contribute to the overall X-ray luminosity in the pulsar.
This is also seen  from the variation of accretion column radius ($r_{0}$) with luminosity. We see an increase of $r_{0}$ value with increasing the source luminosity up to the outburst peak. As the outburst fades. after attaining the peak luminosity, 
the column emission radius also decreases to its pre-outburst peak value $\le500$~m. The dominance of bulk-Comptonization during 
the peak luminosity is evident from the variation of $\delta$ parameter with luminosity. This parameter signifies the 
effective role of thermal Comptonization and bulk Comptonization processes in shaping the pulsar accretion column. 

At low X-ray luminosity, its expected that, the energizing of photons takes place through thermal scattering of 
plasma electrons. Whereas at  high X-ray luminosity, its the bulk motion of electrons energizes the X-ray photons 
in accretion column via inverse Compton scattering. At peak luminosity, we see that, its the pure bulk 
Comptonization process (as indicated by the higher values of $\delta$), that dominates net Comptonized 
column emission from the pulsar. There is an interesting variation of accretion plasma temperature ($kT_{e}$)
with the luminosity. At higher source luminosity, we see a slight decrease in the $kT_{e}$, implying a cooling
of plasma emission at the outburst peak. This may happen, as the most of the hard X-ray photons escape from the 
side walls of accretion column above {\it critical luminosity} during which the emission beam pattern from the 
pulsar switches from a pure pencil beam emission at low X-ray luminosity to a mixture of pencil-beam and 
fan-beam emission pattern after attaining its critical luminosity.  The dimensionless parameter $\xi$ is
expressed in equations (26,103,104) of \cite{becker2007thermal} in terms of ratio of the dynamical time scale 
($t_{shock}$) for the accretion of gas onto the neutron star from the sonic point down to the NS surface to 
the time scale ($t_{esc}$) for the radiation to diffuse through the column walls of the radiation dominated 
accretion column. The ratio of these two timescales relates to $\xi$ (through equation 104 of \citealt{becker2007thermal}) 
as : $\frac{t_{shock}}{t_{esc}}\sim0.24\xi$. Comparing our estimated values of 
$\xi$ for \exo (as shown in Table~\ref{table4}), we find that these two time scales are of comparable 
magnitude. This shows that the accretion flow in the column is radiation dominated where photons
find the least resistance to escape in the side walls of accretion column rather than upward diffusion along the column.
The column emission radius or the hotspot radius ($r_{0}$),  is found to be increasing with source 
luminosity reaching a maximum value $800$ m at the outburst peak for few observation epochs in the high luminosity range of $(\sim 7.8-9.7)\times10^{37}$~\lum. However during most of the outburst, 
the estimated hot spot radius $\leq500$ m is within the theoretical constraints (see equation 112 in \citealt{becker2007thermal}).

\begin{table*}
\centering 
\caption{Best-fitting spectral parameters with $1\sigma$ errors obtained from \hxmt observations of \exo with BW model.}
\resizebox{\textwidth}{!}{
\begin{tabular}{cccccccccccc}
\hline 
& & \multicolumn{6}{c}{BW model Parameters} \\
& & \\\cline{3-8} 
Obs-IDs     &Luminosity$^a$    &$\dot{M}$        &$\xi$    &$\delta$         &B                  &T$_{e}$           &r$_{0}$ &$E_{gabs}$       &$Strength$        &${\sigma_{gabs}}$  &\redchi(d.o.f)\\
            &($10^{37}$~\lum)  &($10^{17}\,\,erg\,\,s^{-1}$) &        &                 &($10^{12}$~G)      &(\kev)            &(m)     &(\kev)           &                 &(\kev)   & \\
\hline \\

P0304030004$^{\ddag}$ &\(4.34 \pm 0.15\) &2.34   &\(2.12 \pm 0.14\) &\(5.81 \pm 0.75\) &\(7.35 \pm 0.72\)  &\(7.51 \pm 0.42\) &480.2   &\(44.7 \pm 0.9\) &\(3.42 \pm 0.44\) &5.0      &1.187(1117) \\
P0304030005 &\(4.91 \pm 0.02\) &2.64   &\(2.08 \pm 0.10\) &\(5.43 \pm 0.49\) &\(7.10 \pm 0.42 \) &\(7.33 \pm 0.28\) &275.2   &\(47.3 \pm 1.0\) &\(2.24 \pm 0.33\) &5.0      &1.110(1236) \\
P0304030008 &\(4.88 \pm 0.21\) &2.63   &\(2.00 \pm 0.11\) &\(5.37 \pm 0.60\) &\(6.87 \pm 0.58 \) &\(7.16 \pm 0.28\) &623.5   &\(48.6 \pm 0.7\) &\(3.78 \pm 0.42\) &5.8      &1.108(1236) \\
P0304030011 &\(6.91 \pm 0.02\) &3.72   &\(4.91 \pm 1.27\) &\(1.27 \pm 0.33\) &\(7.94 \pm 0.22 \) &\(9.48 \pm 0.27\) &317.2   &--               &--                &--       &1.031(1237) \\
P0304030012 &\(6.69 \pm 0.02\) &3.60   &\(3.17 \pm 0.55\) &\(2.30 \pm 0.81\) &\(7.16 \pm 0.70 \) &\(8.39 \pm 0.26\) &363.4   &\(47.3 \pm 1.1\) &\(1.67 \pm 0.29\) &5.0      &1.045(1235) \\
P0304030013 &\(5.68 \pm 0.03\) &3.06   &\(2.50 \pm 0.12\) &\(4.17 \pm 0.45\) &\(8.13 \pm 0.95 \) &\(8.37 \pm 0.45\) &365.6   &\(50.1 \pm 0.6\) &\(2.73 \pm 0.31\) &5.0      &1.053(1235) \\
P0304030016 &\(6.15 \pm 0.23\) &3.31   &\(2.81 \pm 0.18\) &\(2.50 \pm 0.30\) &\(6.14 \pm 0.09 \) &\(7.57 \pm 0.16\) &371.9   &\(54.2 \pm 0.5\) &\(3.62 \pm 0.51\) &5.0      &1.036(1236) \\
P0304030017 &\(6.21 \pm 0.30\) &3.34   &\(8.93 \pm 1.50\) &\(0.61 \pm 0.13\) &\(7.82 \pm 0.03 \) &\(9.85 \pm 0.06\) &364.8   &\(52.1 \pm 1.8\) &\(0.95 \pm 0.01\) &5.0      &1.092(1236) \\
P0304030018 &\(5.82 \pm 0.27\) &3.13   &\(2.52 \pm 0.21\) &\(3.24 \pm 0.53\) &\(6.65 \pm 0.25 \) &\(7.83 \pm 0.21\) &399.9   &\(51.9 \pm 0.6\) &\(3.23 \pm 0.30\) &5.0      &0.998(1237) \\
P0304030020 &\(6.13 \pm 0.18\) &3.30   &\(2.44 \pm 0.13\) &\(3.20 \pm 0.34\) &\(6.52 \pm 0.15 \) &\(7.64 \pm 0.15\) &390.0   &\(49.6 \pm 0.6\) &\(2.57 \pm 0.22\) &5.0      &1.037(1229) \\
P0304030021 &\(6.68 \pm 0.10\) &3.60   &\(1.47 \pm 0.03\) &\(9.59 \pm 0.52\) &\(7.02 \)          &\(4.80 \pm 0.18\) &405.9   &\(55.2 \pm 0.5\) &\(15.88 \pm 0.85\) &11.6    &1.042(1354) \\
P0304030024 &\(7.02 \pm 0.16\) &3.78   &\(1.81 \pm 0.08\) &\(5.87 \pm 0.50\) &\(7.16 \)          &\(6.37 \pm 0.30\) &405.6   &\(44.9 \pm 1.0\) &\(3.25 \pm 0.35\) &5.0      &1.109(1352) \\
P0304030025 &\(7.15 \pm 0.15\) &3.85   &\(2.58 \pm 0.20\) &\(2.83 \pm 0.40\) &\(7.39 \pm 0.37\)  &\(8.02 \pm 0.31\) &465.5   &\(46.4 \pm 1.3\) &\(1.49 \pm 0.30\) &5.0      &1.163(1355) \\
P0404147001 &\(7.22 \pm 0.09\) &3.89   &\(2.18 \pm 0.05\) &\(7.87 \pm 0.40\) &\(6.67 \pm 0.25\)  &\(6.29 \pm 0.10\) &468.8   &\(45.0 \pm 4.7\) &\(0.37 \pm 0.13\) &5.0      &1.124(1347) \\
P0304030026 &\(7.45 \pm 0.15\) &4.01   &\(1.46 \pm 0.03\) &\(9.85 \pm 0.56\) &\(7.74 \)          &\(4.60 \pm 0.17\) &498.2   &\(46.3 \pm 0.5\) &\(5.36 \pm 0.34\) &5.0      &1.077(1351) \\
P0304030027 &\(7.62 \pm 0.12\) &4.10   &\(1.75 \pm 0.03\) &\(16.07 \pm 0.45\) &\(6.00 \)         &\(4.10 \pm 0.12\) &661.0   &\(49.6 \pm 0.7\) &\(2.27 \pm 0.24\) &4.9      &1.002(1345) \\
P0304030029 &\(7.79 \pm 0.11\) &4.19   &\(1.15 \pm 0.01\) &\(20.00 \pm 0.05\) &\(2.88 \)         &\(2.59 \pm 0.13\) &811.0   &\(39.7 \pm 0.4\) &\(10.62 \pm 0.64\) &8.36    &1.104(1345) \\
P0304030030 &\(8.13 \pm 0.22\) &4.38   &\(2.13 \pm 0.04\) &\(13.24 \pm 1.15\) &\(6.05 \pm 0.22\) &\(5.48 \pm 0.21\) &824.2   &--               &--                &--       &1.081(1360) \\
P0304030031 &\(7.95 \pm 0.16\) &4.28   &\(2.08 \pm 0.03\) &\(13.86 \pm 0.72\) &\(5.44 \pm 0.10\) &\(5.08 \pm 0.12\) &800.5   &--               &--                &--       &1.152(1360) \\
P0304030032 &\(8.10 \pm 0.17\) &4.36   &\(2.04 \pm 0.03\) &\(15.20 \pm 0.79\) &\(5.20 \pm 0.09\) &\(4.87 \pm 0.12\) &800.5   &--               &--                &--       &1.021(1353) \\
P0304030033 &\(8.12 \pm 0.13\) &4.37   &\(2.00 \pm 0.02\) &\(17.06 \pm 0.64\) &\(6.35\)          &\(4.68 \pm 0.12\) &800.9   &--               &--                &--       &1.112(1353) \\
P0304030034$^{\ddag}$ &\(9.65 \pm 0.63\) &5.19   &\(2.16 \pm 0.03\) &\(14.51 \pm 0.77\) &\(5.91 \pm 0.12\) &\(5.41 \pm 0.09\) &820.7   &--               &--                &--       &1.143(1346) \\
P0304030035 &\(7.80 \pm 0.11\) &4.20   &\(2.17 \pm 0.04\) &\(10.97 \pm 0.62\) &\(5.80 \pm 0.10\) &\(5.44 \pm 0.12\) &401.3   &--               &--                &--       &1.103(1352) \\
P0304030036 &\(8.33 \pm 0.10\) &4.49   &\(2.00 \pm 0.02\) &\(16.61 \pm 0.41\) &\(5.81 \pm 0.40\) &\(4.61 \pm 0.07\) &393.8   &--               &--                &--       &1.137(1352) \\
P0304030037 &\(8.00 \pm 0.10\) &4.30   &\(2.17 \pm 0.04\) &\(11.74 \pm 0.81\) &\(6.10 \pm 0.22\) &\(5.66 \pm 0.15\) &394.1   &--               &--                &--       &1.003(1353) \\
P0304030038 &\(8.15 \pm 0.12\) &4.39   &\(1.95 \pm 0.04\) &\(15.93 \pm 0.68\) &\(6.43\)          &\(4.39 \pm 0.15\) &401.0   &--               &--                &--       &0.995(1361) \\
P0304030039 &\(7.92 \pm 0.11\) &4.26   &\(1.88 \pm 0.03\) &\(16.70 \pm 0.57\) &\(6.02\)          &\(4.12 \pm 0.12\) &392.2   &--               &--                &--       &1.022(1360) \\
P0304030040 &\(7.41 \pm 0.14\) &3.99   &\(2.17 \pm 0.03\) &\(13.97 \pm 0.68\) &\(6.47 \pm 0.41\) &\(5.45 \pm 0.12\) &394.0   &--               &--                &--       &1.172(1360) \\
P0304030041 &\(6.82 \pm 0.15\) &3.67   &\(2.13 \pm 0.04\) &\(16.65 \pm 0.70\) &\(6.49 \)         &\(4.77 \pm 0.14\) &393.9   &--               &--                &--       &1.141(1349) \\
P0304030042 &\(6.47 \pm 0.16\) &3.48   &\(1.62 \pm 0.11\) &\(7.35 \pm 1.30\)  &\(4.94 \pm 1.28 \) &\(5.28 \pm 0.42\)  &465.0 &\(44.4 \pm 1.0\) &\(4.66 \pm 0.64\) &5.0      &1.112(1110) \\
P0304030044 &\(6.07 \pm 0.17\) &3.27   &\(4.08 \pm 0.88\) &\(1.63 \pm 1.19\)  &\(8.04 \pm 0.35 \) &\(9.59 \pm 0.25\)  &484.4 &\(45.7 \pm 1.0\) &\(1.82 \pm 0.32\) &5.0      &1.032(1025) \\
P0304030046 &\(5.59 \pm 0.15\) &3.01   &\(3.29 \pm 0.22\) &\(3.41 \pm 0.52\)  &\(9.46 \pm 0.73 \) &\(10.17 \pm 0.62\) &359.5 &\(46.1 \pm 1.6\) &\(0.91 \pm 0.23\) &3.9      &1.187(1124) \\
P0304030050$^{\ddag}$ &\(4.04 \pm 0.17\) &2.17   &\(4.49 \pm 1.16\) &\(8.73 \pm 2.49\)  &\(9.21 \pm 0.51 \) &\(13.41 \pm 0.78\) &369.0 &\(42.1 \pm 1.2\) &\(1.11 \pm 0.29\) &5.0      &1.030(1102) \\
P0304030051 &\(3.38 \pm 0.17\) &1.82   &\(4.64 \pm 0.69\) &\(11.15 \pm 1.94\) &\(9.35\)          &\(13.94 \pm 0.65\)  &369.4 &\(44.2 \pm 1.1\) &\(1.39 \pm 0.24\) &5.0      &1.032(1104) \\ \hline

\end{tabular}
}
\flushleft
Notes:\\
$^a$ : The 1-100 \kev luminosity in the units of 10$^{37}$~\lum by assuming a distance of 7.1~\kpc. \\
$^\ddag$: Indicates the representative Obs-IDs from which the unfolded spectra are best-fitted with BW Model and shown in 
Figure~\ref{fig:13}. The computed BW model parameters for these Obs-IDs are shown in Table~\ref{table5}.\\
\label{table4}
\end{table*}

\begin{deluxetable*}{cccccccccc}
\tabletypesize{\scriptsize}
\tablecaption{Computed BW Model Parameters.\label{table5}}
\tablewidth{0pt}
\tablehead{
\colhead{OBSID}
& \colhead{$\alpha$}
& \colhead{$\sigpar/\sig$}
& \colhead{$\sigbar/\sig$}
& \colhead{$J$ (g\,s$^{-1}$\,cm$^{-2}$)}
& \colhead{$\Tmound$ (K)}
& \colhead{$\vmound / c$}
& \colhead{$\taumound$}
& \colhead{$\taumax$}
& \colhead{$\tautrap$}
}
\startdata
P0304030004
&0.594
&$52.53 \times 10^{-5}$
&$12.18 \times 10^{-4}$
&$3.23 \times 10^7$
&$1.56 \times 10^7$
&0.015
&0.026
&0.80
&1.30
\\
P0304030034
&0.605
&$30.05 \times 10^{-5}$
&$3.94 \times 10^{-4}$
&$2.45 \times 10^7$
&$1.50 \times 10^7$
&0.016
&0.027
&0.68
&1.29
\\
P0304030050
&1.257
&$8.04 \times 10^{-5}$
&$1.47 \times 10^{-4}$
&$5.07 \times 10^7$
&$1.80 \times 10^7$
&0.016
&0.013
&0.34
&0.89
\\
\enddata


\end{deluxetable*}

\section{DISCUSSIONS}\label{discuss}
In this paper, we have presented the timing and spectral analysis of \exo using \hxmt observations in 2021 which 
monitored the source between X-ray luminosity range of $\sim 0.6\times 10^{38}$ and $1.1\times 10^{38}$\lum. The obtained pulse profiles evolve with luminosity and energy are shown in Fig~\ref{fig:2} and 
Fig~\ref{fig:3}. The energy dependence of the pulse fraction are presented in Fig~\ref{fig:4}. We found for the first time a clear 
cyclotron line energy at 47\kev (see Fig~\ref{fig:7}). The existence of 47\kev\ CRSF in \exo is not dependent on the continuum spectral models 
(e.g., \hcut\ and \cpl ). This 47 \kev absorption structure can also be observed in the phase-resolved spectrum within 
several phase intervals, as shown in Fig~\ref{fig:9}. The pulse profile evolution and spectral parameter variations with luminosity suggest the spectral transition near the critical luminosity near $(0.8-1)\times 10^{38}$ erg s$^{-1}$. In the followings, we use critical luminosity model to discuss the magnetic 
field strength of the NS.
\par
The transition from subcritical to supercritical accretion regimes around critical luminosity is anticipated to bring 
alterations to the geometry of the emission region, consequently influencing the shape of the pulse profile. This phenomenon
has been discussed in several published articles \citep{becker2012spectral,becker2007thermal,wang2022timing,mushtukov2015critical}.
At lower luminosity, the deceleration of the accretion flow can occur through Coulomb breaking within a plasma cloud. 
The stopping region of the flow is positioned just above the neutron star (NS) surface, and the emission originating 
from this stopping region escapes from the top of the column, forming a pencil beam. At luminosity above $L_{crit}$, 
deceleration is predominantly dominated by radiation pressure, with emission primarily escaping through the column walls, 
forming a fan-beam. The emission pattern transitions from a pencil-beam at low luminosity to a fan-beam at higher luminosity. 
The beam pattern translate from a fan-beam to a pencil-beam which is usually accompanied by a conversion between the two-peak 
pulse profile and the one-peak pulse profile. The strong changes in the pulse profile shape are observed at a luminosity of 
$\sim 0.8\times 10^{38}$\lum.
\par
Parameters such as photon index, column density, and cutoff energy exhibit an inverse correlation with luminosity at 
low luminosity. With the increase in luminosity, there is an inverse correlation compared to low luminosity. 
Due to absence of substantial bulk Comptonization of photons with the accreting electrons, the pulsar spectrum appear 
soft \citep{becker2012spectral,epili2017}. For the subcritical area (luminosity$\leqslant 8\times 10^{37}$\lum), 
the accretion flow still goes through the radiation-dominated shock, and due to the pressure of the radiation is 
insufficient to bring the matter to rest at the stellar surface, the accretion flow will be decelerated by 
Coulomb braking ultimately. The photon index $\Gamma$ and luminosity show negative correlation, due to the emission zone 
decreases with the increasing luminosity. It seems like \exo undergo a transition from a negative to positive evolution 
and show a flat trend during 2021 outburst, in the range between $(0.8-1)\times 10^{38}$\lum. The luminosity of the transition 
revealed by the spectrum coincides with the changes in pulse profiles. Those phenomena seem to conclude that these are two 
distinct transitions that occurred in the range within a certain luminosity range $(0.8-1)\times 10^{38}$\lum. 
The critical luminosity at which the emission mode shifts relies on the strength of the neutron star's magnetic field. 
This can be calculated as follows\citep{becker2012spectral}:

\begin{equation}
\begin{aligned}
L_{\text{crit}} = & 1.49 \times 10^{37} \, \text{erg} \, \text{s}^{-1} \left( \frac{\varLambda}{0.1} \right)^{-\frac{7}{5}} \omega^{-\frac{28}{15}} \\
& \times \left( \frac{M_*}{1.4M_{\odot}} \right)^{\frac{29}{30}} \left( \frac{R_*}{10 \, \text{km}} \right)^{\frac{1}{10}} \left( \frac{B_*}{10^{12} \, \text{G}} \right)^{\frac{16}{15}} .
\end{aligned}
\end{equation}

For the case of \exo, we assume that neutron star mass and radius values $M_*$ = 1.4$M_*$ and $R_*$ = 10km, and 
$\varLambda $ = 0.1 and $\omega$ = 1 based on the theoretical considerations. The inferred magnetic field strength range for the NS in \exo is $\sim (4.8-6.0) \times 10^{12}$G.
\par
Cyclotron resonance scattering features are the only way to directly measure the surface magnetic field strength of a neutron star. Several authors considered that there was a possible cyclotron absorption feature around 10\kev in \exo
\citep{wilson2006detection,klochkov2007integral,wilson2008outbursts}, although the \hxmt observations suggest that such an absorption structure may be model-dependent rather than a genuine characteristic. If this is indeed a cyclotron absorption feature, the corresponding magnetic field is $\sim 1\times 10^{12}$ G. \cite{reig1999x} attributed 
a possible spectral absorption feature at 36\kev to a cyclotron absorption line, the corresponding magnetic field strength 
will be $\sim 3.7\times 10^{12}$G. From a CRSF at $\sim47$\kev\ reported by 
\hxmt, we estimated the NS surface magnetic field strength of 
\exo and gave it to be $\sim 4.9\times 10^{12}$ G, which is good agreement with the magnetic field strength based on the critical luminosity. 

\par
The 2021 giant outburst of \exo shows that, the column emission is radiation dominated around the outburst peak. This allows us to use the physics based BW model \citep{becker2007thermal} to explore the column emission of the pulsar assuming the
canonical values of neutron star and the accretion rate estimated from observed X-ray luminosity of the pulsar during the
outburst. With the BW model applied to the phase averaged spectra of \exo, we are able to estimate the fundamental 
parameters such as : $\delta$, $B$, $r_{0}$ and $T_{e}$. The variations of these parameters with the outburst luminosity 
could illustrate the role of thermal and bulk Comptonization of the accretion plasma in shaping the pulsar column emission
which is radiation dominated. Based on these fundamental parameters, the BW model can also be used to compute several 
additional parameters of the model. 

In Table~\ref{table5}, we show some of these physical parameter values obtained using the BW model. 
These are namely: (1) $\alpha$: the constant of the assumed velocity profile ($v(\tau) = -\alpha c \tau$),
(2) $\sigpar/\sig$: the ratio of the scattering cross section of photons propagating parallel the magnetic field to the 
Thomson scattering, (3) $\sigbar/\sig$: angle averaged cross section in units of the Thomson value, (4) $J$: the mass 
accretion flux, (5) $\Tmound$(K): The temperature of the thermal mound in $cgs$, (6)$\vmound / c$ : The inflow speed at the 
mound surface in terms of $c$, (7) $\taumound$: optical depth at the top of the thermal mound, (8) $\taumax$: The maximum 
optical depth, a dynamical constraint for the assumed velocity profile such that, at large distances from the neutron star, 
the radiation pressure is negligible and (9) $\tautrap$: The optical depth of the trapped radiation in the rapidly falling 
gas in the  accretion column. These  parameters are defined in \citet{becker2007thermal} respectively in the equations 
(33),(83),(84),(92),(93),(88),(89),(79) \& (107). 

From our calculations, we find that, the inflow speed at the thermal 
mound $\vmound\sim0.02c$, whereas the free fall velocity at the top of accretion column is $v_{ff}\sim0.6c$. This shows a 
significant deceleration of the accretion plasma at the NS surface during the outburst. The mound temperature ($\Tmound$(K)) is
found to be increasing as the mass accretion flux ($J$) increases. From the comparison of $\taumax$ and $\tautrap$ values , 
we find that, for all the three luminosity epochs of observations as mentioned Table~\ref{table5}, $\taumax \leqslant 
\tautrap$. This indicates that, the trapped region of the accretion column is from where most of the observed emission is 
produced.  In the radiation dominated accretion column as the matter falls rapidly, the photon ``trapping'' occurs when the 
downward advective flux of photons dominates the upward diffusion of photons along the column axis \citep{becker2007thermal}.
This leads to confinement of photons in the lower regions of the accretion flow. However through diffusion, the radiation 
is effectively transported vertically from the lower regions of the column. We find that in addition to the observed 
luminosity epochs as mentioned in Table~\ref{table5} for \exo, at the other epochs of the 2021 outbursts also 
(as noted in Table~\ref{table4}), the condition $\taumax \leqslant \tautrap$ holds good. This is a further indication that, 
during the giant outburst, as the accretion column is highly radiation dominated, the bulk Comptonization traps the 
radiation in the lower regions of accretion column. 

Apart from the bulk Comptonization being an efficient mechanism of energy transfer from the accreting gas 
to the the photons, we find that, in \exo during the progress and declining of the giant outburst, the thermal 
Comptonization as well plays a vital role in formation of broadband spectra. The phase-averaged spectra \exo has been 
successfully explained with quasi-exponential cutoff at high energies (i.e \hcut) with a flattening of spectrum at low 
energies. The contribution of these spectral shapes is mainly due to thermal processes that transfer the energy from high 
to low frequency radiation. Now this is evident from application of BW model, from which we can see the thermal 
Comptonization playing a significant role via Compton scattering of high energy photons. The flattened spectra is due to 
the subsequent inverse Compton scattering  of soft photons by the recoiling plasma electrons leading to energy transfer to 
low frequency photons \citep{becker2007thermal}.

\section{Summary}
Using high-cadence and high-statistic observations by \hxmt, we investigate the timing and spectral properties of 
this source during the brightest type II outburst of \exo occurring in 2021 in details. A previously unnoticed cyclotron 
absorption line is reported in the spectrum. In the pulse phase-averaged spectra, the fundamental CRSF is clearly detected 
between 44 and 50\kev\ with the different continuum models. Phase-resolved spectral analyses consistently show strong evidence 
of CRSF around 47\kev. A transition from sub-critical to super-critical regime is seen in the variations of spectral parameters, 
which will be due to changes in the emission geometry across the critical luminosity.
\par
There is an energy dependence and a luminosity dependence of the pulse profile shape. The rms pulse fraction drops sharply 
at 30\kev\ appearing near luminosity $1.1\times10^{38}$\lum. The pulse profile exhibits significant changes between 
two peaks and one peak around the luminosity corresponding to the transition between subcritical and supercritical 
accretion regimes. Based on the calculation of the critical luminosity model, the magnetic field of \exo can be estimated 
to be $\sim\left(4.8- 6.0\right)\times10^{12}$ G. In addition, the BW model is applied to the wide band spectrum, 
and constrains the pulsar magnetic field in the range of $\left(5-9 \right) \times 10^{12}$ G.


\section*{Acknowledgements}
We are grateful to the referee for the suggestions to improve the manuscript. This work is supported by the NSFC (No. 12133007) and the National Key Research and Development Program of China 
(Grants No. 2021YFA0718503). This work has made use of data from the \textit{Insight}-HXMT mission, a project funded 
by the China National Space Administration (CNSA) and Chinese Academy of Sciences (CAS).



\bibliography{sample631}{}
\bibliographystyle{aasjournal}



\end{document}